\documentclass[oldversion]{aa}
\usepackage{graphicx}
\usepackage{txfonts}
\usepackage{natbib}
\usepackage{lscape}
\bibpunct{(}{)}{;}{a}{}{,}

\begin{document}

\title{Estimation of stellar atmospheric parameters from SDSS/SEGUE spectra}
\author{P. Re Fiorentin \inst{1}
        \and
        C.A.L. Bailer-Jones \inst{1}
        \and
        Y.S. Lee \inst{2}
        \and
        T.C. Beers \inst{2}
        \and 
        T. Sivarani \inst{2}
        \and
        R. Wilhelm \inst{3}
        \and
        C. Allende Prieto \inst{4} 
        \and 
        J.E. Norris \inst{5}       
        }
\offprints{P. Re Fiorentin, \email{fiorent@mpia.de}}

\institute{Max Planck Institut f\"ur Astronomie, K\"onigstuhl 17, 69117
  Heidelberg, Germany. 
        \and
        Department of Physics \& Astronomy, CSCE: Center for the Study of
        Cosmic Evolution, and JINA: Joint Institute for Nuclear Astrophysics,
        Michigan State University, East Lansing, MI 48824, USA 
        \and 
        Department of Physics, Texas Tech University, Lubbock, TX 79409, USA
        \and
        Department of Astronomy, University of Texas, Austin, TX 78712, USA
        \and
        Research School of Astronomy and Astrophysics, Australian National
        University, Weston, ACT 2611, Australia
}

\date{Received 21 February, 2007; accepted Day March, 2007}

\abstract{We present techniques for the estimation of stellar atmospheric
  parameters ($T_{\rm eff}$, $\log~g$, ${\rm [Fe/H]}$) for stars from the
  SDSS/SEGUE survey. The atmospheric parameters 
  are derived from the observed medium-resolution ($R = 2000$) stellar spectra
  using non-linear regression models trained either on (1) pre-classified
  observed data or (2) synthetic stellar spectra. In the first case we use our
  models to automate and generalize parametrization produced by a preliminary
  version of the SDSS/SEGUE Spectroscopic Parameter Pipeline (SSPP). In the
  second case we directly model the mapping between synthetic spectra (derived
  from Kurucz  model atmospheres) and the atmospheric parameters, 
  independently of any intermediate 
  estimates.  After training, we apply our models to various samples of SDSS
  spectra to derive atmospheric parameters, 
  and compare our results with those obtained previously by the SSPP for the
  same samples. We obtain consistency between the two approaches, with RMS
  deviations on the order of $150$~K in $T_{\rm eff}$, $0.35$~dex in $\log~g$,
  and $0.22$~dex in ${\rm [Fe/H]}$. 
  
  The models are applied to pre-processed spectra, either via Principal
  Components Analysis (PCA) or a Wavelength Range Selection (WRS) method, which
  employs a subset of the full 3850--9000\,$\AA$ spectral range.  This is both
  for computational reasons (robustness and speed), and because it delivers
  higher accuracy (better generalization of what the models have learned).
  Broadly speaking, the PCA is demonstrated to deliver more accurate 
  atmospheric parameters when the training data are the actual SDSS
  spectra with previously estimated parameters, whereas WRS 
  appears superior for the estimation of $\log~g$ via synthetic templates,
  especially for lower signal-to-noise spectra. 
  From a subsample of some 19\,000 stars with previous determinations of 
  the atmospheric parameters accuracies of our predictions (mean absolute
  errors) for each 
  parameter are $T_{\rm eff}$ to $170/170$~K, $\log~g$ to
  $0.36/0.45$~dex, and ${\rm [Fe/H]}$ to $0.19/0.26$~dex, for methods (1) and
  (2), respectively.  We measure the intrinsic errors of our models by training
  on synthetic spectra and evaluating their performance on an independent set
  of synthetic spectra. This yields RMS accuracies of $50$~K, $0.02$~dex, and
  $0.03$~dex on $T_{\rm eff}$, $\log~g$, and ${\rm [Fe/H]}$, respectively.  
  
  Our approach can be readily deployed in an automated analysis pipeline,
  and can easily be retrained as improved stellar models and synthetic spectra
  become available. We nonetheless emphasise that this approach relies on an
  accurate calibration and pre-processing of the data (to minimize mismatch
  between the real and synthetic data), as well as sensible choices concerning
  feature selection.

From an analysis of cluster candidates with available SDSS spectroscopy
(${\rm M~15}$, ${\rm  M~13}$, ${\rm M~2}$, and ${\rm NGC~2420}$), and assuming
the age, metallicity, and distances given in the literature are correct, we
find evidence for small systematic offsets in $T_{\rm eff}$ and/or $\log~g$
for the 
parameter estimates from the model trained on real data with the SSPP. 
Thus, this model turns out to derive more precise, but less accurate, 
atmospheric parameters than the model trained on synthetic data.

\keywords{Astronomical data bases: Surveys -- Methods: data analysis --
  Methods: statistical -- Stars: fundamental parameters -- Galaxy:
  globular/open clusters individual: ${\rm M~15}$, ${\rm  M~13}$, ${\rm
    M~2}$/${\rm NGC~2420}$}} 

\maketitle

\section{Introduction}\label{introduction}

The nature of the stellar populations of the Milky Way galaxy remains an
important issue for astrophysics, because it addresses the question of galaxy
formation and evolution and the evolution of the chemical elements. To date,
however, studies of the stellar populations, kinematics, and chemical
abundances in the Galaxy have mostly been limited by small number statistics.

Fortunately, this state of affairs is rapidly changing. The Sloan Digital Sky
Survey \citep[SDSS; ][]{york} has imaged over $8000$ square degrees of the
northern Galactic cap (above $|b| = 40^o$) in the $ugriz$ photometric system
for some 100 million stars.  
Imaging data are produced simultaneously 
\citep{fukugita, gunn98, gunn06, hogg, dr3,dr5} and processed through
pipelines to measure photometric and astrometric properties \citep{lupton,
  stoughton, smith, tucker, pier, ivezic} and to select targets for
spectroscopic follow-up. 
Of even greater importance, some 200\,000 medium-resolution stellar spectra
have been obtained during the course of SDSS-I (the original survey).

The SDSS-II project, which includes SEGUE (Sloan Extension for Galactic
Understanding and Exploration), is obtaining some 3500 square degrees of
additional imaging data at lower Galactic latitudes, in order to better explore
the interface between the thick-disk and halo populations between $0.5-4$~kpc
from the Galactic plane. 
SEGUE will obtain some 250\,000 medium-resolution spectra of stars in the
Galaxy in the magnitude range $14.0 \le g \le 20.5$ in 200 fields covering the
sky visible from the northern hemisphere (Apache Point Observatory, New
Mexico). The targets are selected based on the photometry, and are chosen to
provide tracers of the structure, chemical evolution, and stellar content of
the Milky Way from 0.5 to 100 kpc from the Sun. Taken together, the stellar
database from SDSS-I and SEGUE provides an unprecendented opportunity for
developing better understanding of the properties of the Milky Way.

Of special importance to achieve these goals is the determination of intrinsic
stellar physical properties, such as masses, ages, and elemental
abundances. The first step toward achieving this goal is to estimate the
atmospheric parameters 
for these stars. A number of early studies
\citep[e.g., ][]{gulati, cbj97, cbj98, cbj00, snider, willemsen} have
demonstrated that non-linear regression models can be robust and precise
classifiers of stellar spectra, either when trained on pre-classified observed
data or on synthetic stellar spectra. In this paper we further explore the
capability of these techniques to estimate $T_{\rm eff}$, $\log~g$, and ${\rm
[Fe/H]}$ specifically for SDSS/SEGUE spectroscopy and photometry. Alternative
procedures are described by \citet{carlos}, \citet{lee06}, and
\citet{lee07}.

In this paper we explore three approaches in which either synthetic (`S') or
real ('R') data are used for training and/or testing. With {\bf SS} (training
and testing on synthetic data), estimates of the 
atmospheric parameters are obtained from the model spectra, and the
application is merely a test of the limits of the pre-processing/regression
model. In {\bf RR} (training and testing on real data), we use a set of
pre-parametrized SEGUE spectra, in this case from a preliminary version of the
SDSS/SEGUE Spectroscopic Parameter Pipeline (SSPP). 
Our model automates and, more importantly, generalizes these parametrizations.
The model performance is evaluated on a separate set of data obtained from
SDSS/SEGUE. {\bf SR} is a model trained on synthetic data and applied to real
data, thus allowing us to directly determine 
the atmospheric parameters without using an intermediate
parametrization model. As we have no definitive ``true'' 
values against which to
compare our parametrizations, we instead compare the results of the SR and RR
models to 
parameters estimated by the SSPP (on a set of data not used to train
RR). Of course, in both the SR and RR cases the derived 
parameters are based on a set
of model atmospheres -- the difference is how 
the atmospheric parameters are derived from them.

The layout of this paper is as follows. In Sect. 2 we describe the
spectroscopic and photometric data from which preliminary estimates of the 
atmospheric parameters were obtained. Our regression model is described
in Sect.~\ref{model}. In Sect.~\ref{dimensionality} we discuss the advantages
of dimensionality reduction via Principal Component Analysis, as well as from
wavelength (``feature'') selection. The results of the application of our
methods using the SS, RR, and SR approaches are discussed in
Sect.~\ref{results}. 
An independent assessment of the accuracy (and calibration) of our models is
provided in Sect.~\ref{cluster}, where we estimate the 
atmospheric parameters of stars in several
Galactic globular and open clusters. Finally, in Sect.~\ref{concluding} we
provide our conclusions. 


\section{Data}
In this section we discuss the SDSS/SEGUE spectra and the synthetic spectra that
were constructed in order to build our models.

\begin{figure*}
   \centering
   \includegraphics[angle=-90,width=17cm]{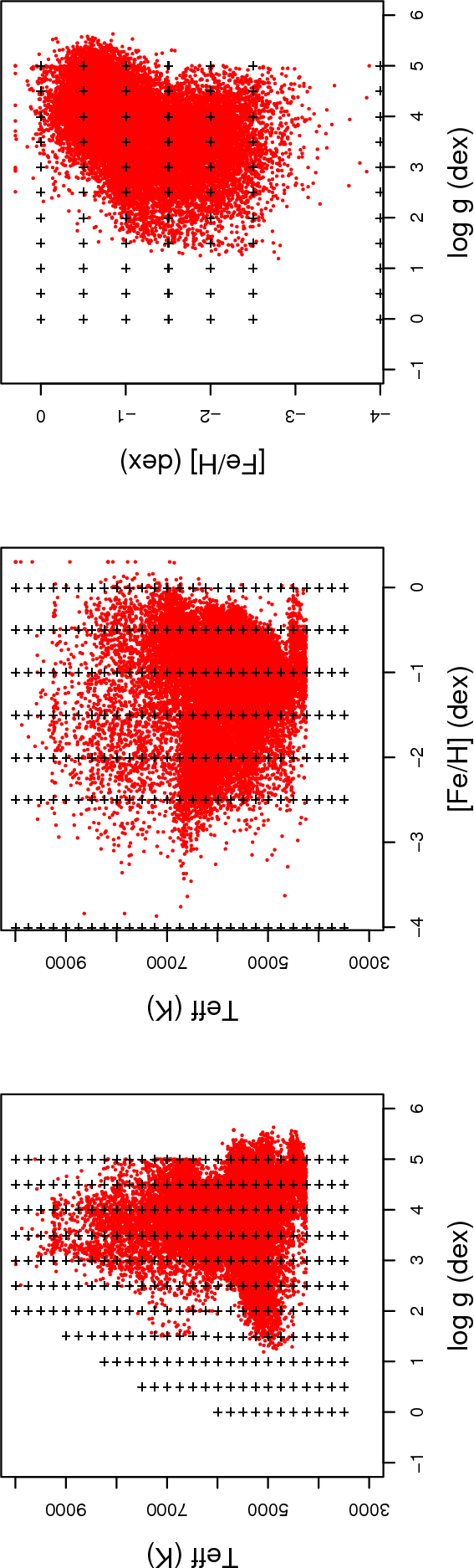}
   \caption{The grid of stellar atmospheric parameters $T_{\rm eff}$,
     $\log~g$, and ${\rm [Fe/H]}$. The synthetic parameters (plus symbols) are
     presented in comparison with previously estimated atmospheric parameters
     (black dots) for 38\,731 SDSS/SEGUE spectra. } 
\label{grid}
\end{figure*}

\subsection{Sample of real spectra}\label{sample_real}

Stellar spectra from SDSS/SEGUE cover the wavelength range 3850--9000\,$\AA$
at a resolving power $R = \lambda/\Delta\lambda\simeq 2000.$ The spectra are
wavelength calibrated and approximately flux corrected using procedures
described in \citet{stoughton}.  
For the purpose of our work, we first rebin to 
  a final dispersion of 1.0\,${\bf\AA}$/pixel in the blue region
  3850--6000\,$\AA$, and 1.5\,${\bf\AA}$/pixel in the red region
  6000--9000\,$\AA$. 
Since the spectrophotometric corrections applied to these spectra are only
approximate, we remove the continuum via an automated, iterative procedure
(described in Sect.~\ref{sample_synthetic}).

We have selected a sample of 38\,731 stellar spectra for stars in regions
of low reddening, and for which 
atmospheric parameter estimates of 
effective temperature, gravity, and metallicity ($T_{\rm eff}$, $\log~g$, ${\rm [Fe/H]}$)
have been obtained previously using the combination of 
procedures described in the SSPP \citep{lee07}, including several
that rely on the available $ugriz$ photometry. 
These methods include chi-square minimization with respect to synthetic
  spectral templates, 
  neural networks, autocorrelation analysis, and a variety of line index
  calculations based on previous calibrations with respect to known standard
  stars. Estimates of the likely external errors in spectroscopic
  parameter determinations are in the process of being obtained by comparison
  with a number of previously available stellar spectroscopic libraries, as
  well as with high-resolution spectroscopy of over 100 SDSS/SEGUE stars.  The
  use of multiple methods allows for empirical determinations of the internal
  errors for each parameter. 
  However, we remark that at present the parameters from SSPP are
  inhomogeneously assembled, in the sense that we are still in the process of
  exploring which techniques are optimal over the parameter ranges which we
  study. This situation will change in the near future, when the techniques
  involved in the SSPP can be evaluated more fully, and are used to 
  produce a meaningful weighted average. 

Radial velocities estimated by the SSPP are
used to reduce all spectra to a common radial velocity zero point. 

\subsection{Sample of synthetic spectra}\label{sample_synthetic}

In recent years a number of new atmospheric
models covering a wide range of atmospheric parameters have become
available. Here we make use of a set of 1816 synthetic spectra calculated from
Kurucz's NEWODF models \citep{castelli_2003} with solar abundances by
\citet{asplund}, including ${\rm H_2O}$ opacities, an improved
set of ${\bf \rm TiO}$ lines, and no convective overshoot 
\citep{castelli_1997}. 
All pertinent molecular species are included in these models, 
even those whose features have minor strength in the wavelength range
  covered by the SDSS spectra.  
The synthetic spectra are generated using the {\tt turbospectrum} synthesis
code \citep{alvarez}, and employ line broadening according to the prescription
of \citet{barklem98}. The linelists used come from a variety of sources.
Updated atomic lines are taken mainly from the VALD database \citep{kupka}. 
The molecular species CH, CN, and OH are provided by B. Plez 
\citep[see ][]{plez}, while the NH, ${\bf \rm C_2}$ molecules are from the
Kurucz linelists (see http: //kurucz.harvard.edu/LINELISTS/LINESMOL/).
Note that, at present, the linelists used to generate the synthetic spectra do
not include all of the interesting molecular species, in particular, the MgH
and CaH features. We plan to include these molecules in an updated version of
our synthetic spectra, which is now under construction.

Our grids span the parameter ranges [3500, 10000]~K in $T_{\rm eff}$ (27
values, stepsize of $250$~K), [0, 5] in $\log~g$ (11 values in 0.5 dex steps),
and [-4.0, 0.0] in ${\rm [Fe/H]}$ (7 values, stepsize between $0.5$ dex and
$1.5$ dex; there is gap in the grid between ${\rm [Fe/H]} = -2.5$ and $-4.0$).
The synthetic spectra are similarly divided into blue and red regions, and the
same dispersion correction and flux ``calibration'' (i.e.\ instrument
modeling) were applied to match the real SDSS/SEGUE spectra. Figure~\ref{grid}
shows the grid of the available parameters. The data used cover the full input
range provided, 3850--9000\,$\AA$, in $4152$ individual data bins. It
should also be noted that we have not implemented any procedure to account for
the inevitable presence of telluric lines, in particular near the location of
the calcium triplet. At present, new reductions procedures for SDSS spectra are
being explored to minimize the impact of telluric lines in this region. 

The continuum is removed by dividing the spectrum by an iterative fifth-order
polynomial fit of the spectrum.  This is done separately for the blue and red
regions. In the following we exclude the red region 6000--6500\,$\AA$, because
we found that the synthetic spectra do not properly model the real ones. 
This discrepancy may be due in part to instrumental signatures 
in this spectral region, which corresponds to the wavelengths where 
the dichroic used in the dual-arm SDSS spectrographs split the incoming
photons into the blue and red arms.

\section{Non-linear regression model}\label{model}

We implement a flexible method of regression that provides a global
non-linear mapping between a set of inputs (the stellar spectrum 
${\rm \bf x}_i$) and a set of outputs (the stellar 
atmospheric parameters, 
${\bf s}=\{T_{\rm eff}, \log~g, {\rm [Fe/H]}\}$)

\begin{equation}
{\bf s}(p)=f\left(\sum_i{w_{i}{\rm \bf x}_{ip}}\right)
\end{equation}

\noindent where $p$ denotes the $p^{th}$ flux vector (star) and $w_{i}$ the
set of weights that characterise the regression model \citep{cbj00}.
To reduce the dynamical range of $T_{\rm eff}$ and to better represent the
uncertainties we use $\log~T_{\rm eff}$. Furthermore, in order
to put all variables ($\bf s$ and $\rm {\bf x}_{\it ip}$) on an equal footing,
we set, for each variable, the mean to zero and standard deviation to unity (a
linear conversion). This helps with the internal stability of most machine
learning algorithms.

The free parameters, \{$w$\}, of the model are the learned error minimization
using sets of data for which inputs and their corresponding outputs are known.
This is an iterative procedure in which patterns are presented to the model,
the outputs calculated, and the difference between these and the target outputs
are used to perturb the weights in a direction that reduces the error.  Learning
is stopped once the rate of reduction of the error drops below some threshold.
Our error function comprises two parts. The first term in the equation below
is the sum-of-squares error in the predictions (the likelihood), the second is
a regularization term,
\begin{equation}
E = \sum_p\left(\frac{1}{2} \sum_l {\beta_l[ y(p)_l - T(p)_l ]^2}\right) +
\alpha \frac 12\sum_i{w_{i}^2} 
\end{equation}
where, for each pattern $p$, $T(p)_l$ and $y(p)_l$ are the target value and its
estimate from the regression method for the $l^{th}$ 
atmospheric parameter, respectively. The
model (hyper)parameters $\beta_l$ dictate the relative importance of each 
parameter in the total error, and $\alpha$ specifies the degree of
regularization. In the present work these hyperparameters were optimized via a
brute force search (conditioned by experience). We actually use a
``committee'' of ten identical models trained on the same data, but trained
from different initial random weights.  
Estimates of the atmospheric parameters obtained by the application of
the model are the average of the ten individual estimates. This simple
approach helps overcome ``convergence'' noise and, on average, increases the
accuracy obtained. 

Our estimate of the accuracy of the model in the application phase is the mean
absolute error
\begin{equation}
E=\frac 1N \sum_{p=1}^N{|C(p)-T(p)|}
\end{equation}
where $C(p)$ is the committee estimate, and $T(p)$ is an independent estimate
for the $p^{th}$ spectrum.  For the SR and RR models (see Sect.~\ref{results})
$T$ is an estimate based on other methods (e.g., SSPP), so $E$ is not a real
``error'', but rather a discrepancy (as there is no definitive ``ground
truth''). 

\begin{figure*}
   \centering
   \includegraphics[angle=0,width=17cm]{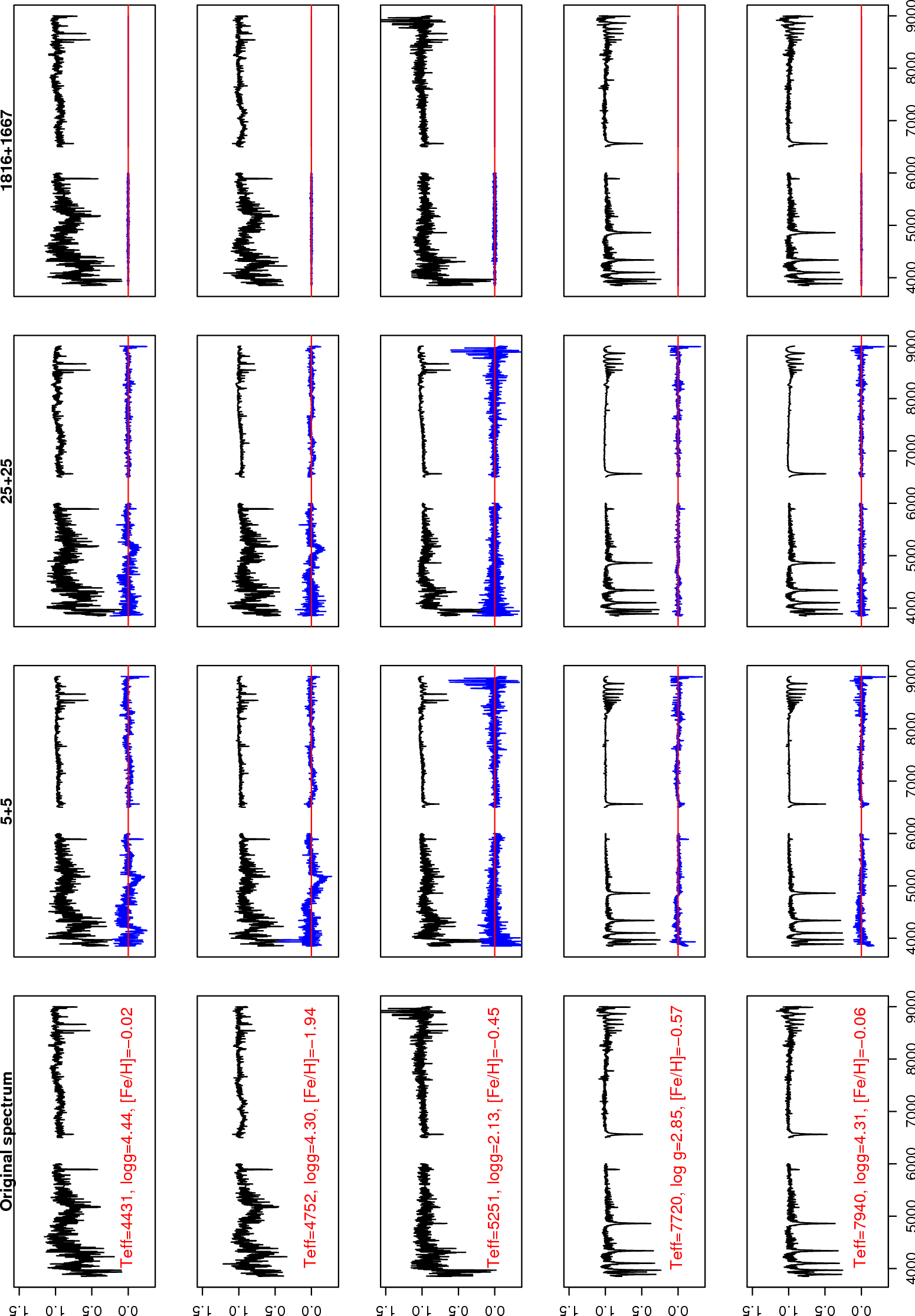}
   \caption{Reconstruction of SDSS/SEGUE spectra by projection onto 
     synthetic principal components. In each row, the spectrum on the left is
     the original and the following show the reconstruction using increasing
     numbers of principal components.  The residual spectrum (original minus
     reconstructed) is shown in the bottom of each panel.  The
     quoted atmospheric parameters are taken from a preliminary version of
     the processing pipeline SSPP.}
\label{recS}
\end{figure*}

\begin{figure*}
   \centering
   \includegraphics[angle=0,width=17cm]{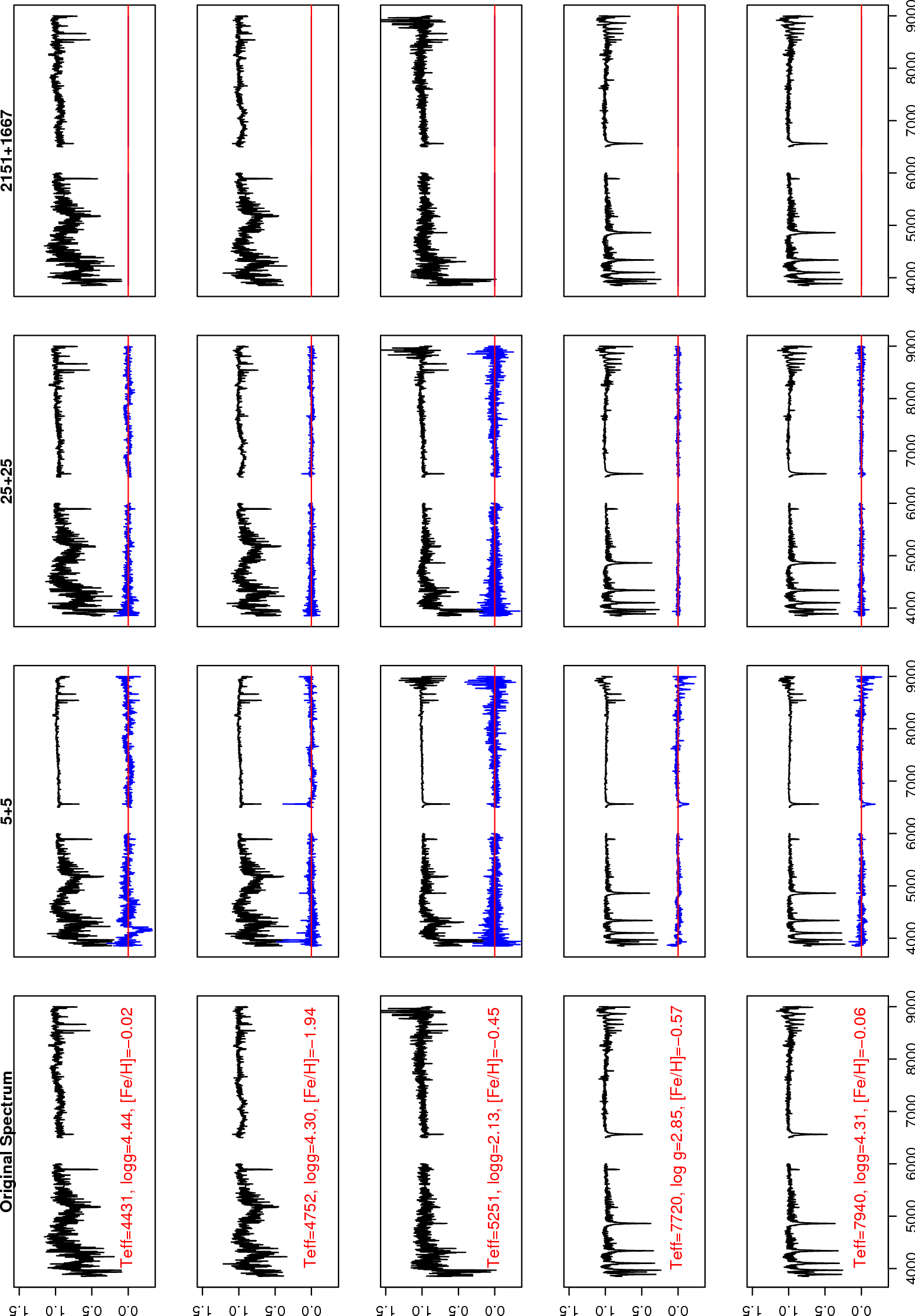}
   \caption{As Fig.~\ref{recS} but for principal components built from real
     spectra.}
   \label{recR}
\end{figure*}
\section{Dimensionality reduction}\label{dimensionality}

Our initial models based on the full spectrum produced good results, but we
find that the full spectrum is not necessary (not surprisingly, as it contains
a large amount of redundant information).  Dimensionality reduction often
leads to enhanced reliability, because of the smaller number of parameters
employed, and the considerably reduced computing time.  We investigated
various approaches and retained two -- Principal Component Analysis
\citep[e.g.; ][and references therein]{ESL, singh, cbj98} and a Wavelength
Range Selection \citep[e.g., ][]{beers99, willemsen} -- in the present work.

\subsection{Principal Component Analysis (PCA)}\label{pca}

Principal Component Analysis (PCA) linearly transforms a set of data via a
rotation of the coordinate system, and an offset of its origin. The new axes
(or principal components, the PCs) are chosen such that the projection of the
data onto each axis in turn maximizes the variance in the data. 
If we have a set of $n$ vectors (spectra), $x$, of dimension $N$ (the number
of flux bins), then formally the principal components are the eigenvectors,
$\mathbf{u}_k$ ($k=1 \ldots N$), of the covariance matrix of the data.
The $p^{th}$ spectrum is reconstructed using the PC basis as
\begin{equation}
\mathbf{y}_p(r)=\sum_{k=1}^{k=r}{a_{kp} \mathbf{u}_k}
\end{equation}
where
\begin{equation}
a_{kp}=\mathbf{x}_p\cdot \mathbf{u}_k
\end{equation}
are the so-called ``admixture coefficients''. These represent the spectrum in
the new (PC) space in the same way that the original spectrum did in the
original (flux bin) space (i.e., they can be used as inputs in our regression
models). If we set $r=N$ then we reconstruct the spectra exactly. If $r<N$ we
have a reduced reconstruction, i.e., a compression which uses just the $r$ most
significant PCs (those with the largest eigenvalues). 

If the number of spectra is smaller than the dimensionality of the data, i.e., 
if $n < N$, then the spectra span a subspace of dimensionality $n$. In this
case only $n$ PCs are defined and a full reconstruction is achieved with
$N=n$. With $n \geq N$, then using all PCs in the reconstruction means that
{\em any} spectrum -- even one not used to form the PCs -- can be
reconstructed exactly. With $n < N$ this is no longer true.  This is actually
the case with our synthetic data, where $n=1816$ and $N=3818$. This
potentially reduces the quality of any reconstruction, because some of the
data space is not spanned by the PCs.

Reduced spectral reconstructions for five representative SDSS/SEGUE stars,
using different numbers of eigenvectors computed from the synthetic and real
spectra, are shown in Figs. ~\ref{recS} and \ref{recR} respectively. The
residual spectrum, defined as the difference between the original and the
reconstructed spectrum, is shown at the bottom of each column for each pattern
and each reconstruction.  From inspection of these samples, one
can see how the PCA approach acts as an effective
filter to remove noise, recover missing and/or borderline features, and to
detect outliers in a spectrum that are reconstructed with large errors
\citep[e.g., ][]{storrie, cbj98}.
However, here we also note that there is evidence that the Kurucz model
  spectra we have adopted do not well describe SDSS/SEGUE spectra of cool
  stars ($T_{\rm eff} < 5000$~K), especially when a small number of PCs is
  assumed: the residual spectrum of 
  main sequence stars at $T_{\rm eff}= 4431$~K and at $T_{\rm eff}= 4752$~K
  highlights difficulties in reconstructing, with $5+5$ and $25+25$ PCs, 
  the ${\rm C_2}$ band at
  5165\,$\AA$ (see Fig.~\ref{recS}). 

\begin{figure}
   \centering
   \resizebox{\hsize}{!}
   {\includegraphics[angle=90,width=\textwidth]{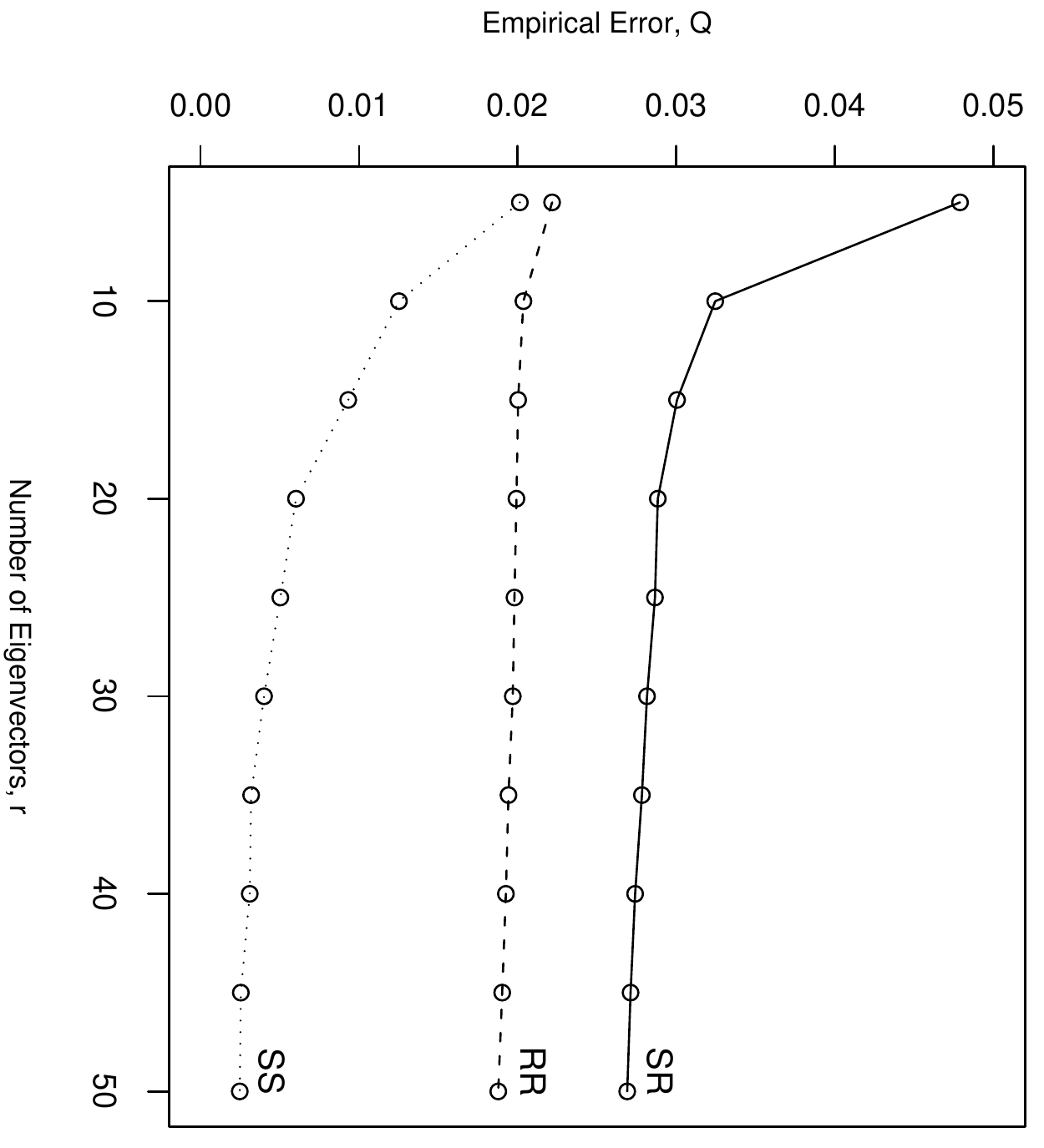}}
   \caption{PCA spectral reconstruction error, $Q$ (defined in
     equation~\ref{pca_rec_err}) on the evaluation data set for SR/RR/SS
     (solid/dashed/dotted lines, respectively) as a function of the
     number of eigenvectors, $r$, used for reconstruction.}
   \label{RQ}
\end{figure}

A useful measure of the reconstruction error over a set of $P$ spectra is
\begin{equation}
Q(r)=\frac{1}{P} \left( \sum_{p=1}^{p=P} \frac{1}{N} \sum_{i=1}^{i=N} \vert
  \mathbf{x}_i - \mathbf{y}^r_i \vert \right) 
\label{pca_rec_err}
\end{equation}
Figure~\ref{RQ} shows how this error varies with $r$. Note that while the PCs
themselves are constructed using the training data set, $Q(r)$ is calculated
on a different set (namely the set to which the regression model is later
applied).  The three cases show quite different behaviour.  For SS the error
drops quite rapidly with increasing $r$, dropping to a constant (but non-zero)
gradient after about 50 PCs (from a total of 1816), whereas for SR and RR the
gradient of the curve becomes constant after including just a few PCs.  The
main reason is that real spectra (used either to form the PCs or in the
projection) show much more variance than synthetic spectra, and this is spread
over more data dimensions. A second observation is that the larger the noise,
the larger the reconstruction error at a given $r$. 
For further discussion see \citet{cbj96} or \citet{cbj98}. 
It is interesting, however, that the curve for SR ``levels off'' at such a low
value of $r$.  This may well be a result of the fact, mentioned above, that
the PCs only span a subspace of the original data space. 

In summary, a PCA compression retains those spectral features which are most
common across the data set. It preferentially removes noise (and rare
features), because they are statistically uncorrelated. Note that the 
atmospheric parameters are not used in defining the PCs. 

Thus, considering the above, the choice of the optimal number of PCs to retain
is a trade-off between retaining information versus reducing dimensionality and
noise, and should be optimized in conjunction with the regression model. There
exist more sophisticated methods of dimensionality reduction which could be
used in the future, such as local and nonlinear variations on PCA 
(see \citet{einbeck} for a review and astronomical application).

\subsection{Wavelength Range Selection (WRS)}\label{wrs}

The restriction of an analysis to certain wavelength intervals via the
exclusion of (hopefully) unimportant ranges, is an alternative way to reduce
the dimensionality of the input space.  This provides a way of directly
introducing domain information into the regression model. While this selection
is potentially difficult (and the number of permutations extremely large), we
show below that this approach is particularly effective for the estimation of
the surface gravity parameter, $\log~g$.  After considering a number of
alternatives, we chose to restrict the analysis on the wavelength ranges
3900--4400\,$\AA$, 4820--5000~\AA, 5155--5350\,$\AA$, and 8500--8700\,$\AA$ in
the spectra.  These regions contain the most prominent hydrogen and metal
lines, including CaII K and H, the Balmer lines H$_\delta$, H$_\gamma$, and
H$_\beta$, the CH G-band, the Mg I{\it b} triplet, and the CaII triplet.


\section{Results}\label{results}

In this section we report the results of the three types of models developed,
SS, RR and SR (for a definition of these see Sect.~\ref{introduction}).

\subsection{SS -- Synthetic vs.\ Synthetic}

For this analysis we adopt the sample of $1816$ noise free synthetic spectra
described in Sect.~\ref{sample_synthetic}. This is randomly split into two
equal-sized sets -- one for model training, and one for model evaluation.

After a preliminary analysis with the full spectra, we decided to use a PCA
pre-processing of the data (Sect.~\ref{pca}).  Principal Components are
computed using the training set, then both sets are projected onto them to
yield the admixture coefficients, which are then the regression model inputs.
PCA is performed on the red and blue spectra separately, because this gave a
better reconstruction (which in turn reduced systematic offsets in the derived
parameters). 
Table~\ref{t_SS} shows typical parametrization errors for the three stellar
atmospheric parameters
for different numbers of PCs retained in the reconstruction; they all are
very small and surprisingly lower for $\log ~g$ than for ${\rm [Fe/H]}$. 
We remark that, when increasing the number of PCs, the error is initially
  determined predominantly by the amount of information present in the
  reconstructed spectra, then by the limited ability of the non-linear
  regression model to make full use of the available information. 
These results, and the analysis of the reconstructed spectra, led us to select
25 (blue region) + 25 (red region) PCs for the model. 

\begin{table}[htb]
  \caption{Mean absolute errors on the evaluation set of 908 spectra in the SS
    model for different numbers of PCs retained in the reconstruction. (As PCA
    is done separately on the red and blue regions, the total number of inputs 
    is twice the number of PCs.)}
  \label{t_SS}
  \begin{center}
    \leavevmode
        \begin{tabular}[h]{llll}
        \hline\hline \\[-5pt]
        PCs& $E_{\log ~T_{\rm eff}}$&$E_{\log ~g}$& $E_{\rm [Fe/H]}$ \\[+5pt]
        \hline \\[-5pt]
        5    &   0.0087 & 0.1264 & 0.1558 \\
        25   &   0.0036 & 0.0245 & 0.0327 \\
        100  &   0.0030 & 0.0251 & 0.0269 \\
        908  &   0.0133 & 0.2087 & 0.2308 \\[+5pt]
       \hline \\[-5pt]
       \end{tabular}
  \end{center}
\end{table}

The above results were obtained with noise-free data, which is not very
realistic, so we also trained models where both the training and evaluation set
are degraded with Gaussian additive noise to signal-to-noise (SNR) levels of
10/1, 30/1, 50/1 and 100/1. Even at a SNR of 10/1, the errors are increased 
{\bf by only} $50$~K in $\log ~T_{\rm eff}$, $0.02$~dex in $\log ~g$, and
$0.03$~dex in ${\rm [Fe/H]}$. 
This modest deterioration is on account of the artifically good correspondence
between the training and evaluation set when using purely synthetic data;
the PCA noise filtering also appears to help. Note that whenever we
use synthetic spectra to define the PCs, we always use noise-free spectra
(also in Sect.~\ref{SR_section}). 

\subsection{RR -- Real vs.\ Real}

Following from our experience with the SS analysis, we build an RR regression
model to parametrize real spectra. The training and evaluation data sets are
taken from a set of 38\,731 stars from 140 SDSS/SEGUE plates, in directions of
low reddening, which have had 
atmospheric parameters estimated by a preliminary version of the
SSPP. Both training and evaluation sets are drawn at random (without
replacement) with sizes 19\,731 and 19\,000 spectra respectively. We use 2151
pixels in the blue spectrum between 3850--6000\,$\AA$ and 1667 pixels in the red
spectrum between 6500--9000\,$\AA$. A PCA compression reduces this to 25 (blue)
$+$ 25 (red) PCs, the PCs themselves formed only from the training set. This
compresses the data to 1.3\% of its former size, resulting in more stable and
faster models. We use these data to predict $\log~T_{\rm eff}$, $\log~g$, and
${\rm [Fe/H]}$. The standard deviations (essentially an estimate of their
parameter ranges) of the input 
parameter distributions are $T_{\rm eff}=724$~K, $\log~g
= 0.64$~dex, and ${\rm [Fe/H]=0.54}$~dex, respectively. These are on the order
of the RMS errors which a random classifier would achieve.

In addition to this purely spectral model, we developed another model in which
the four (de-reddened) photometric colours $u-g$, $g-r$, $r-i$, and $i-z$ are
added as four additional model inputs (they are not involved in the PCA).

\begin{figure*}
   \centering
   \includegraphics[angle=-90,width=17cm]{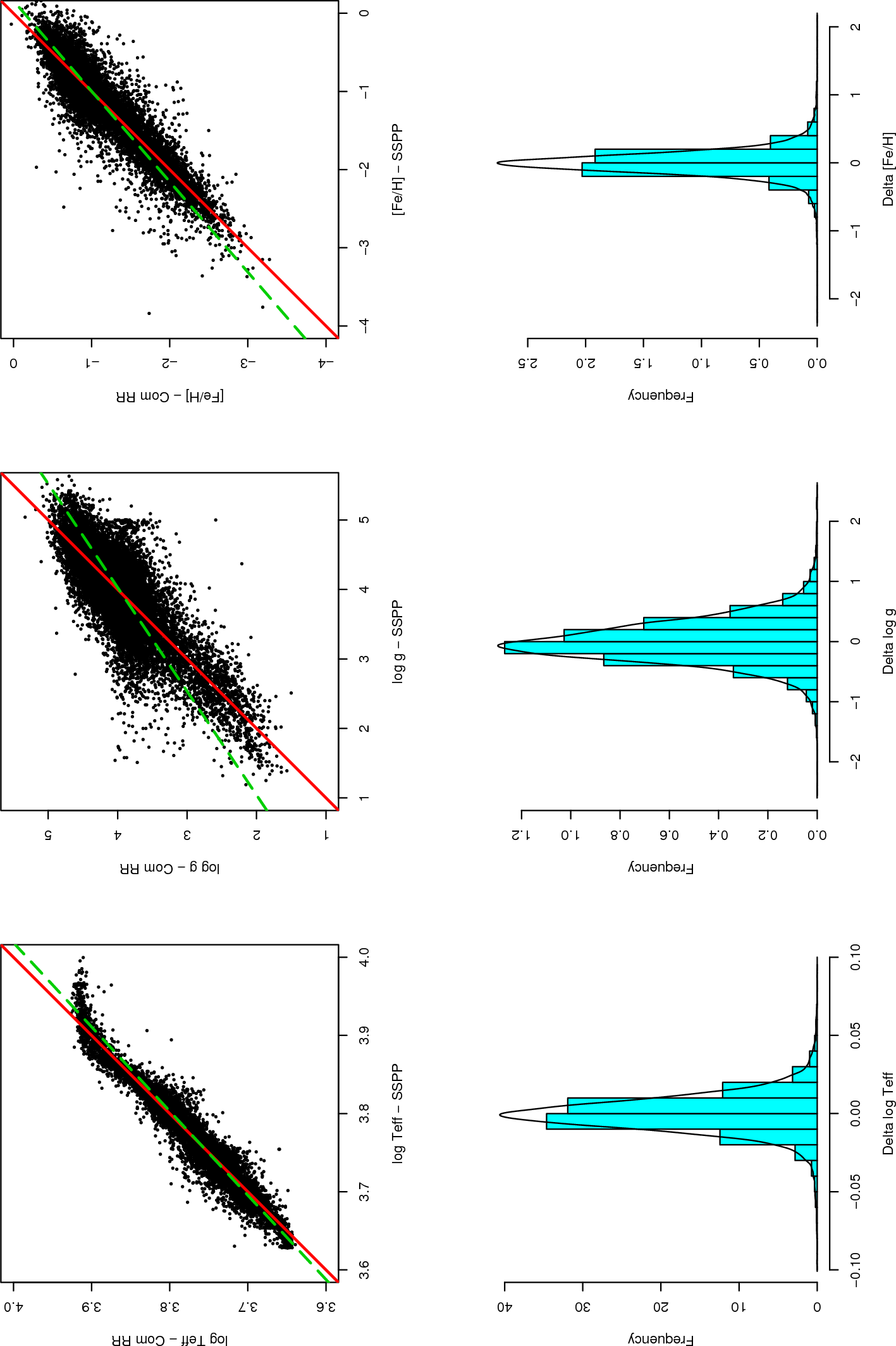}
   \caption{Atmospheric parameters estimation with
     the RR model. We compare our estimated $\log~T_{\rm eff}$, $\log
       ~g$, ${\rm[Fe/H]}$ 
     with those from a preliminary version of the SSPP on the 19\,000 stars in
     the evaluation set. The perfect correlation and a linear fit to the data
     are shown with the solid and dashed lines respectively. The histogram of
     the discrepancies (our estimates minus SSPP estimates) are shown in the
     lower panels. }
   \label{phot}
\end{figure*}

Figure~\ref{phot} compares our model estimates with those from the SSPP on the
evaluation set. Overall we see good consistency, especially for stars with
$T_{\rm eff}<8000$~K ($\log~T_{\rm eff} = 3.90$).  Above this effective
temperature we see that our models underestimate $\log~T_{\rm eff}$ relative
to the SSPP.  Our regression models are designed to smooth, i.e.\ interpolate,
data. Extrapolation of the model to estimate atmospheric parameters 
that are not spanned by the
training set is relatively unconstrained (and any model would need to make
additional assumptions).  Furthermore, the accuracy of the RR model is limited
by the accuracy of the target atmospheric parameters 
used in training, as well as their consistency across the 
parameter space. In this case, the SSPP estimates are
combinations from several estimation models, each of which operates only over a
limited 
parameter range. Thus, the transition we see above 8000\,K may indicate a
temperature region where one of the SSPP submodels is dominating the SSPP
estimates, and this is not well-generalized by our model. Of course, if we
decided that we wanted to reproduce the SSPP predictions for hot stars, we
could do this simply by fitting a second-order polynomial to our residuals to
remove the systematic offset. 


Table~\ref{t_GlobalRR} quantifies the overall discrepacies for each 
parameter.  An
error in $\log~T_{\rm eff}$ of 0.0126 is an error of 2.9\%, or 170\,K at
6000\,K.  The last line in the table is the performance when we include
photometry.  Adding photometry leads to significant improvement in all three
atmospheric parameters.  This is not surprising for effective
temperature, as the photometric calibration of these bands is less complicated
than the spectral calibration. 
A more accurate $T_{\rm eff}$ will permit more accurate $\log~g$ and 
${\rm [Fe/H]}$. Thus, in directions where interstellar redenning is known to
be low, photometry should be used.  The values listed in the table for a given
parameter are averaged over all values of the 
adopted atmospheric parameters.  Results for 
gravity, metallicity, and effective temperature ranges
 -- dwarfs/giants, low/high metallicity, and cool/warm stars -- are listed
in Table~\ref{t_Partial1RR} and in Table~\ref{t_Partial2RR}.

\begin{table}[htb]
  \caption{Mean absolute errors on the evaluation set of 19\,000 spectra in
    the RR model (plotted in Fig.~\ref{phot}). The first line is for the full
    data set (training and evaluation data). The second and third are just for
    the evaluation sets. The third line is for a model which included the four
    photometric colours as additional model inputs (predictors).}
  \label{t_GlobalRR}
\begin{center}
    \leavevmode
        \begin{tabular}[h]{l|lllll}
        \hline\hline \\[-5pt]
        set & PCs & + & $E_{\log~T_{\rm eff}}$ & $E_{\log~g}$& $E_{\rm
          [Fe/H]}$ \\[+5pt]  
        \hline \\[-5pt]
        38\,731 & 25+25  &    & 0.0090 & 0.2699 & 0.1339\\[+5pt]
        \hline \\[-5pt]
        19\,000 &25+25   &    & 0.0126 & 0.3644 & 0.1949\\
                &25+25 &phot& 0.0082 & 0.2791 & 0.1616\\[+5pt]
       \hline
       \end{tabular}
  \end{center}
\end{table}

\subsection{SR -- Synthetic vs.\ Real}\label{SR_section}

We have shown above that our regression models are capable of obtaining
accurate and consistent estimates of 
atmospheric parameters when trained and tested on synthetic
spectra (SS), and also when trained on real spectra with existing parametrizations
and applied to another sample of real spectra (RR).  We now develop the hybrid
approach, SR, in which we train on synthetic spectra and use this model to
determine atmospheric parameters 
for SDSS/SEGUE spectra directly. A very important aspect of this
model is processing the synthetic and real data to look similar;
inaccurate synthetic spectra (e.g.\ poor models or a poor flux calibration)
will degrade performance and/or give rise to systematic errors.

\begin{figure}
   \centering
   \resizebox{\hsize}{!}
   {\includegraphics[angle=90,width=\textwidth]{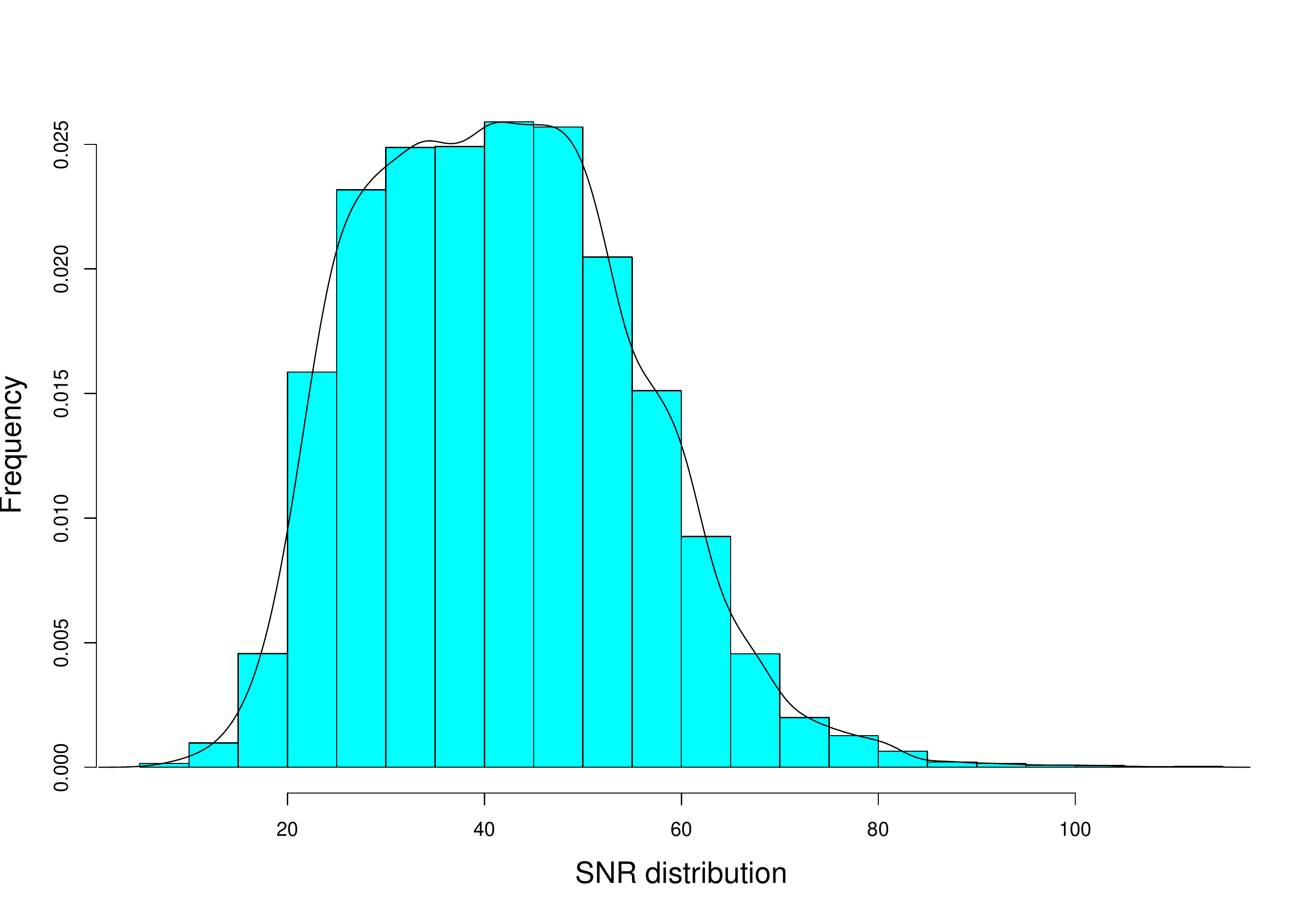}}
   \caption{Histogram of the SNR distribution for all 38\,731 stars of the real
   sample. For each of them, the value for SNR has been estimated from the
   stellar spectrum.}
   \label{SNR_distr}
\end{figure}

Experience shows that it is advantageous to match the noise properties of
the synthetic training sample to that of the real sample. Essentially, noise
acts as a regularizer in the training phase and thus improves the overall
generalization performance of the models \citep[e.g.; ][]{snider, odewahn},
in particular reducing systematics.  For each of the $38\,731$ SDSS/SEGUE
stars in the evaluation set we use the SNR reported (for each pixel) in
the data array included in the FITS file (which was estimated by the reduction
pipeline).  We assign a global SNR to the spectrum which is the median of all
flux bins over the wavelength range we retain (viz.\ 4000--5850\,$\AA$ and
6500--8500\,$\AA$).  Figure~\ref{SNR_distr} shows the distribution of these SNR
values.  Based on this, we chose to develop two regression models, one
optimized for low SNR real spectra (SNR$<$35/1, 13\,487 stars) the other for
high SNR real spectra (SNR$>$35/1, 25\,244 stars).  Experimentation showed that
this noise injection does indeed reduce systematics which are obtained when
using noise-free data for training.

We explored the application of dimensionality reduction with PCA, but found
that this led to rather large systematic errors in the parameters, in
particular in $\log~g$ (up to $1.0$~dex).  We instead found that it is better
simply to select wavelength regions which are known to be the most sensitive
to surface gravity (e.g.\ 3900--4400\,$\AA$, 4820--5000\,$\AA$,
5155--5350\,$\AA$ and 8500-8700\,$\AA$). This is perhaps not unexpected, since
essentially all of the methods that are used by the SSPP to define the target
$\log ~g$ values use only these restricted wavelength ranges.  This may also
indicate that the gravity signature in real stars outside of the wavelength
regions selected above behaves differently from the signature in the synthetic
spectra. Either way, the excluded regions show less sensitivity to $\log ~g$,
so for this parameter these regions do not add information, only 
data that are uncorrelated with the parameter of interest (so are effectively
just noise). It is also possible, of course, that the PCA may be filtering out
subtle (weak) features which are strong predictors of $\log ~g$.

\begin{figure*}
   \centering
   \includegraphics[angle=-90,width=17cm]{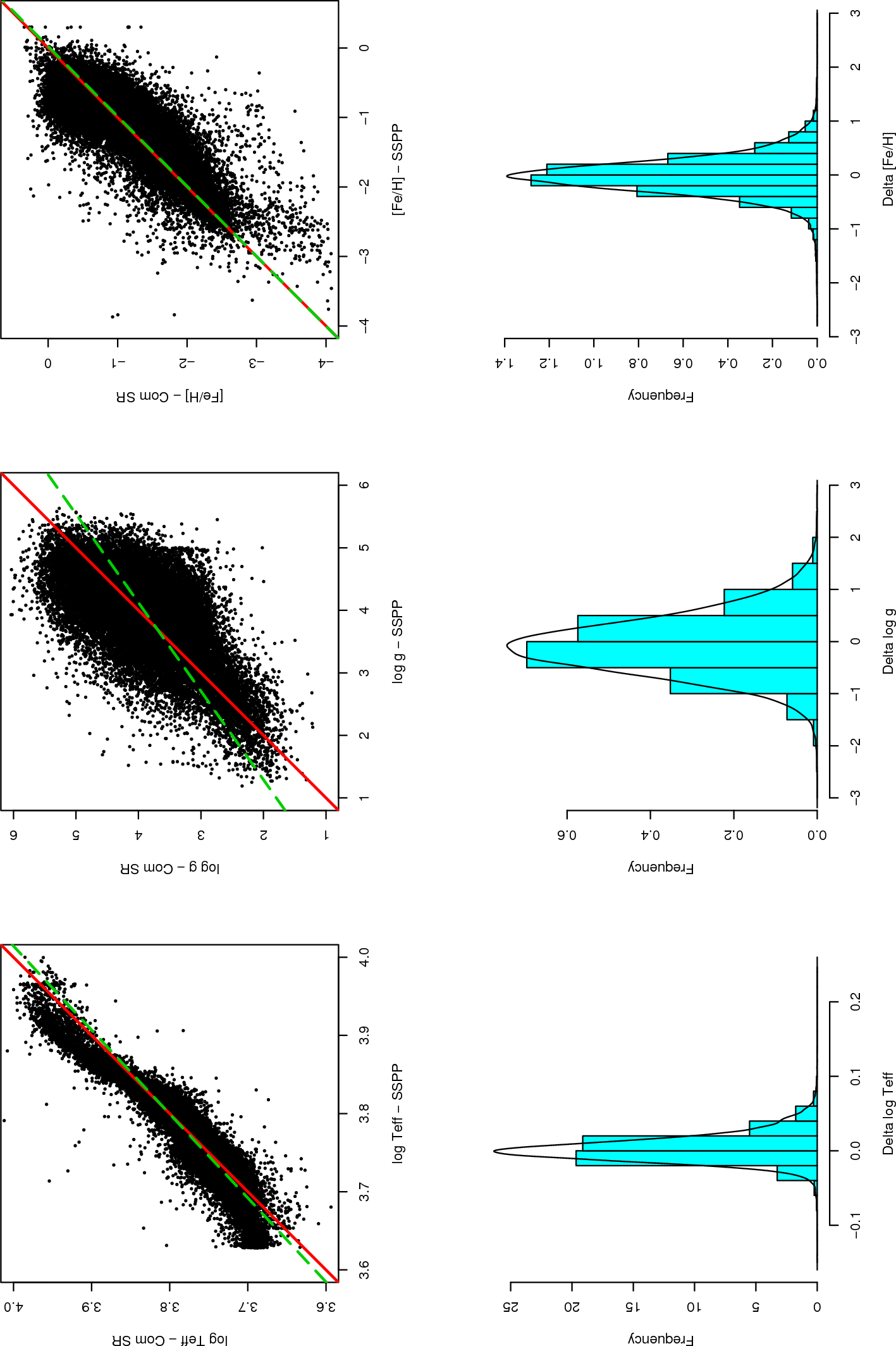}
   \caption{Atmospheric parameters estimation with
     the SR model. Comparison between our derived $\log~T_{\rm eff}$, $\log
       ~g$, ${\rm[Fe/H]}$ and those estimated by a preliminary version of
     SSPP for a set of 38\,731 stars. The perfect correlation and a linear fit
     to the data are shown with the solid and dashed lines respectively. The
     distribution of the residuals (model minus SSPP) are shown in the bottom
     panels. }
   \label{SR}
\end{figure*}

Based on the above considerations, our final model uses PCA for estimating
$T_{\rm eff}$ and ${\rm [Fe/H]}$ and WRS for estimating $\log ~g$.  A separate model
is used for estimating each 
parameter (although the ${\rm[Fe/H]}$ model also predicts the
other two, 
the results of which are disregarded).

Figure~\ref{SR} compares our model 
atmospheric parameter estimates with those from the
preliminary SSPP for the 38\,731 stars in the evaluation set.  While the
overall consistency between the two models is reasonably good, we (again)
notice discrepancies at the extreme 
parameter values, in particular for $T_{\rm eff}$.  
This is sometimes an indication that the model has not been well
trained, i.e., it has not located a good local minimum of the error function
(it can never be shown that the global minimum has been found with anything
but an exhaustive search).  However, there are inevitably problems with
spectral mismatch, in the sense that the synthetic spectra do not reproduce
all of the complexities of the spectra of real stars.  The absence of several
molecular species in the linelists for the synthetic spectra may also be
contributing to this problem, especially for cooler stars where they are
expected to be more important.
For the determination of metallicity, we observe that our model
predicts lower metallicities for the lowest metallicity stars. This is probably
a consequence of the lack of synthetic samples between $-4.0 <{\rm [Fe/H]}<
-2.5$ (see Fig.~\ref{grid}) in our current grid.

\begin{table}[htb]
  \caption{Mean absolute discrepancies (between our SR model and SSPP)
    calculated on the evaluation set of 38\,731 real spectra (see also
    Fig.~\ref{SR}). Our models use PCA pre-processing for estimating ${\rm
      [Fe/H]}$ and $\log~T_{\rm eff}$ and WRS pre-processing for
     estimating  $\log~g$; for the latter, PCA results are shown for
     comparison. 
     Separate models were applied for low and high SNR spectra
    (the transition being at SNR=35/1).
  }
  \label{t_GlobalSR}
  \begin{center}
    \leavevmode
        \begin{tabular}[h]{llccc}
        \hline\hline \\[-5pt]
        Method& SNR&$E_{\log ~T_{\rm eff}}$ &$E_{\log ~g}$&$E_{\rm [Fe/H]}$ \\[+5pt]
        \hline \\[-5pt]
        PCA (25+25) &  & {\bf 0.0138} & 0.4288 & {\bf 0.2606} \\
        &low &0.0143 &0.7549&0.3023\\
        &high&0.0136 &0.3465&0.2384\\[+5pt]
        \hline\\[-5pt]
        WRS&& - & {\bf0.4459}& -\\
        &low &-&0.4495&-\\
        &high&-&0.4450&-\\[+5pt]    
       \hline
       \end{tabular}
  \end{center}
\end{table}

\begin{figure*}
   \centering
   \includegraphics[angle=-90,width=17cm]{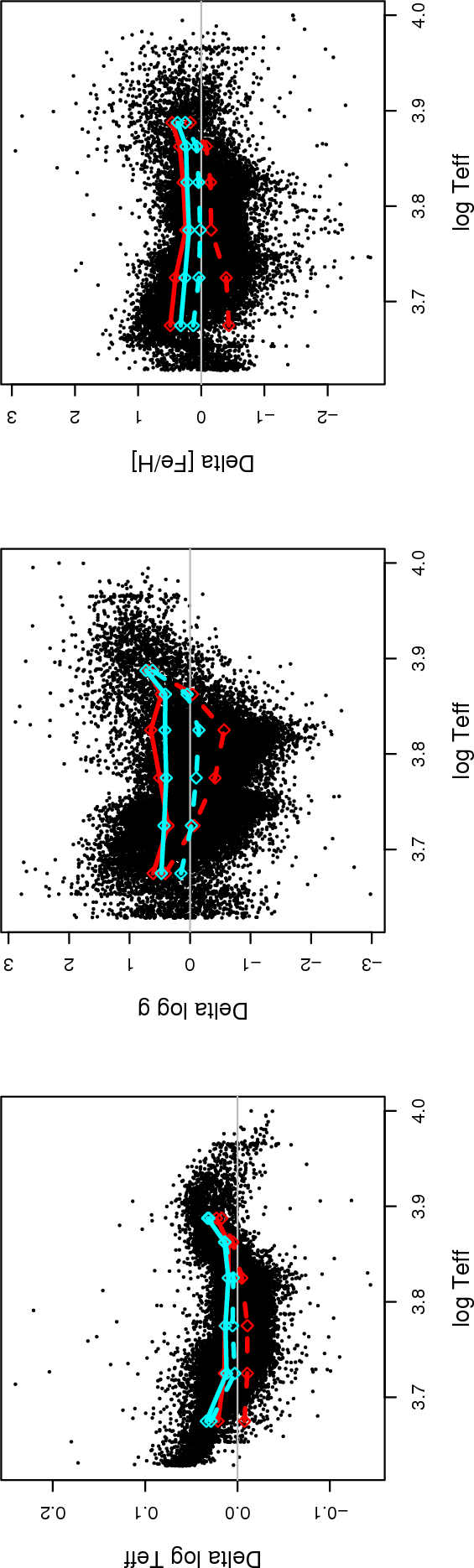}
   \caption{More detailed visualization of the SR model discrepancies
     (Fig.~\ref{SR}).  The diamonds joined by lines show mean absolute
     residual (solid lines) and mean residual (dashed lines) for low
     metallicity (${\rm [Fe/H]}< -1.5$, white lines) and high metallicity
     (${\rm [Fe/H]}> -1.5$, grey lines) averaged over all stars in a bin which
     has the diamond point as its centre. The mean residual traces the
     systematic offset (bias), the mean absolute the scatter.}
   \label{SR_ann_res}
\end{figure*}

Table~\ref{t_GlobalSR} shows the global results (averaged over all stars and 
atmospheric parameters). It is interesting that the WRS
pre-processing results in little difference in the $\log~g$ discrepancy 
for the low and high SNR regimes.
Results for 
gravity, metallicity, and effective temperature ranges
-- dwarfs/giants, low/high metallicity, and
cool/warm stars -- are listed in Table~\ref{t_Partial1SR} and in
Table~\ref{t_Partial2SR}, and visualized in Fig.~\ref{SR_ann_res}. We note
that, in the estimation of $\log~g$, a systematic difference (our model
predictions lower than SSPP) occurs in the range $T_{\rm eff}=5600-6700$~K for
low-metallicity giants. 
Unfortunately we cannot include photometry in the SR models, because the
synthetic colours are not yet well-calibrated, and their zero points on the AB
system are still under discussion. 

\begin{figure*}
   \centering
   \includegraphics[angle=-90,width=17cm]{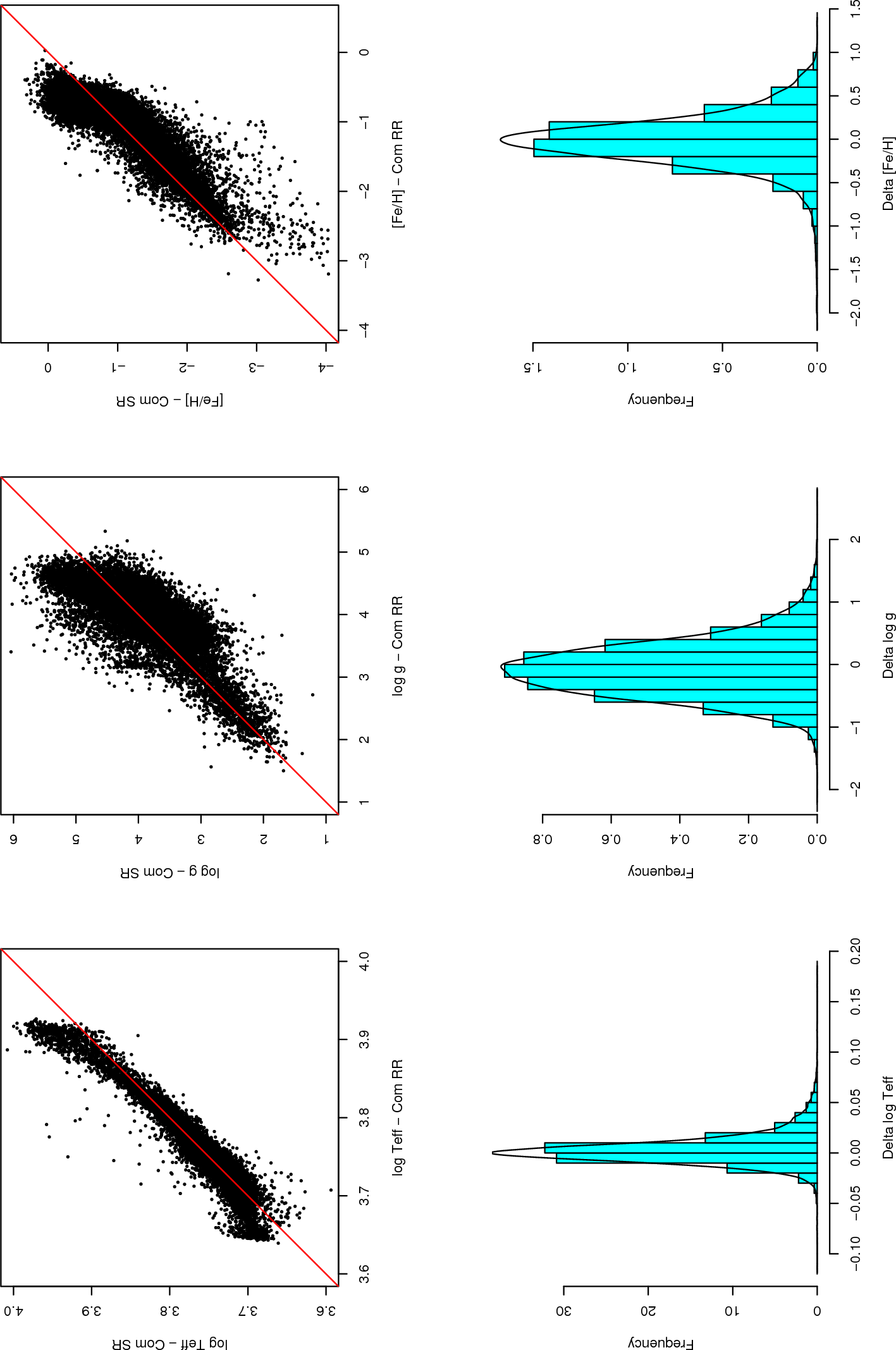}
   \caption{Comparison between SR and RR estimations on the $19\,000$ real
     spectra in common in their evaluation sets. The line shows the perfect
     correlation and the bottom panels the distributions of residuals.}
   \label{SRvRR}
\end{figure*}

\subsection{Comparison of RR and SR}

The RR and SR models developed above both appear to give reasonable
predictions, as measured by their mean accuracies with respect to the SSPP
predictions -- $T_{\rm eff}$ with residual of $0.013/0.014$ ($\sim 170$~K),
$\log~g$ with a residual of $0.36/0.45$~dex and ${\rm [Fe/H]}$ with a residual
of $0.19/0.26$~dex for RR/SR respectively. 

The global discrepancies are larger with SR for $\log~g$ and ${\rm [Fe/H]}$,
but this is not surprising because it is entirely independent of the SSPP 
parameter
estimates.  While the synthetic spectra place a limit on the performance of
the SR model, this is true of any parametrization model. Physical parameters
can only be derived using physical models; none can be measured ``directly''.
The advantage of the SR approach is that it only uses one set in the
parametrizations, it can easily be retrained using new synthetic spectra, and
it provides a quick, general model which operates over the entire 
parameter range. In
effect, the work in getting good predictions is taken out of the machine
learning model and moved to the definition of the templates and the
pre-processing.

We find that PCA delivers more accurate 
atmospheric parameters when the training data are the
actual SDSS spectra with previously estimated parameters, whereas WRS appears
superior for the estimation of $\log~g$ templates, especially from lower SNR
spectra.

From the subsample of $19\,000$ stars used as the evaluation set in RR we
compare the SR predictions with the RR predictions (Fig.~\ref{SRvRR}). The mean
absolute differences are on the order of 0.010 in log $T_{\rm eff}$ ($150$~K),
$0.35$~dex in $\log~g$, and $0.22$~dex in ${\rm [Fe/H]}$.




\section{Application: Globular Clusters}\label{cluster}

Comparison of theoretical isochrones with data from clusters offers an
excellent opportunity to test the present model predictions.  In particular,
we can use them to assess the calibration of the 
parameter determinations.
Here we focus our discussion on the globular cluster ${\rm M~15}$, but we have
also analysed the globular clusters ${\rm M~13}$ and ${\rm M~2}$ and the open
cluster ${\rm NGC~2420}$, all observed by SDSS/SEGUE. We select likely members,
then estimate their 
atmospheric parameters, and overplot these on a set of isochrones fixed
at literature values for the cluster distance modulus, age, and metallicity. If
these values (and the isochrones themselves) are correct, discrepancies between
our 
estimates and the isochrones would indicate problems in the calibrations
of the 
atmospheric parameters (e.g.\ of the synthetic spectra on which the
regression models are based). We note that \citet{lee07} have looked more
carefully at the three globular clusters, and make an independent target
selection based also on stellar densities, from which they derive mean
metallicities and 
radial velocities for the clusters.

\subsection{${\rm M~15}$ - Selection}\label{sect_GCselection}
\begin{figure}
   \centering
   \resizebox{\hsize}{!}
   {\includegraphics[angle=-0,width=\textwidth]{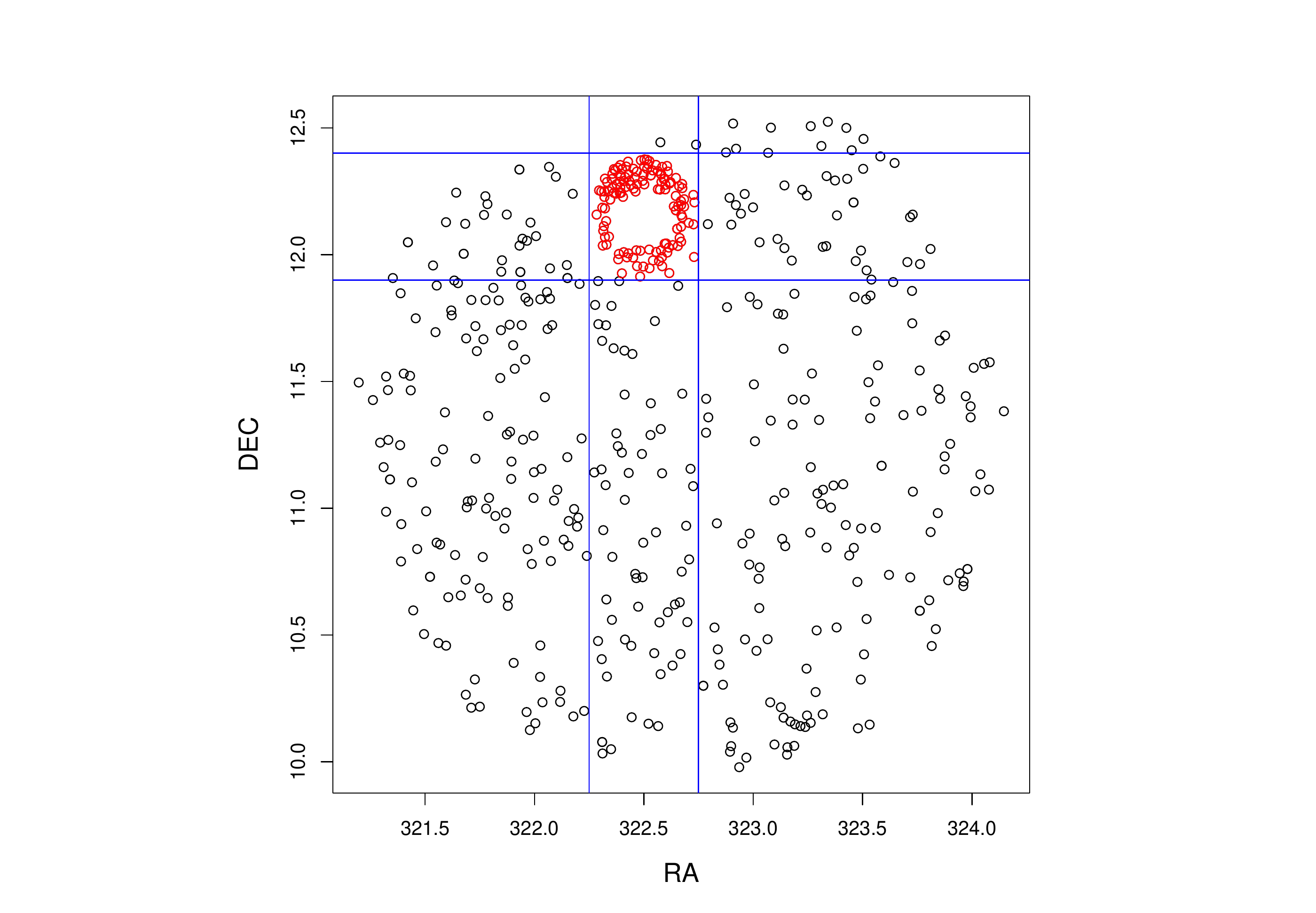}} 
   \caption{Distribution on the sky of the 526
     stars present from SDSS/SEGUE plates 1960 and 1962. The box defines the
     selection criteria ($322^\circ. 25 < \alpha < 322^\circ. 75$ and
     $11^\circ. 90 < \delta < 12^\circ. 40$) which produces 133 ${\rm M~15}$
     candidates.}
   \label{m15_sel}
\end{figure}

\begin{figure}
   \centering
   \resizebox{\hsize}{!}
   {\includegraphics[angle=90,width=\textwidth]{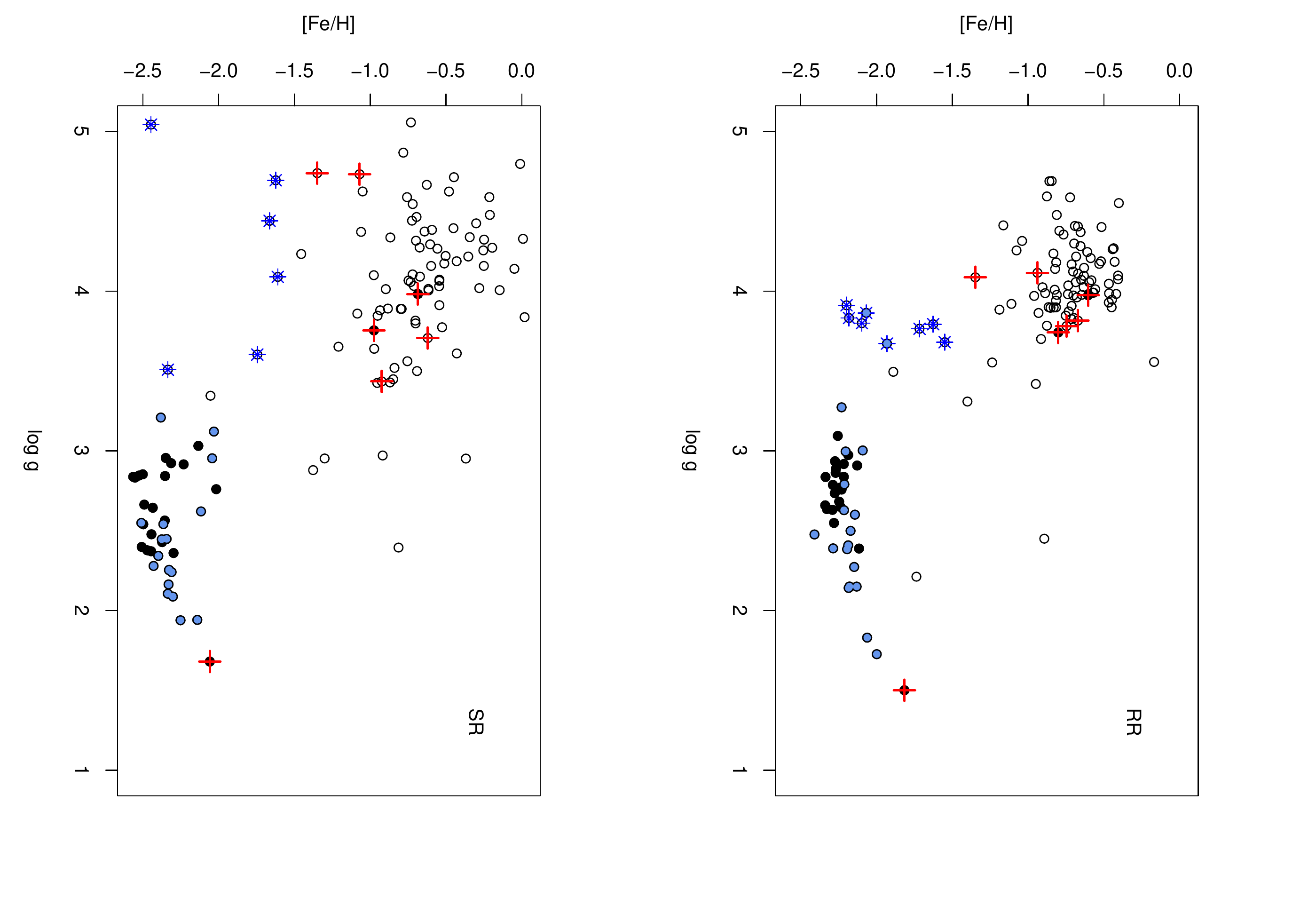}}
   \caption{Distribution of ${\rm [Fe/H]}$ vs. $\log~g$ for the 133
     positionally-selected ${\rm M~15}$ candidates from
     Fig.~\ref{m15_sel}. 
     atmospheric parameters are from the RR (top) and SR (bottom)
     models. Of these 133 candidates, we retain only 46 (RR) or 45 (SR) stars
     in the low-metallicity group as likely cluster members. Among these, 8
     (RR) or 7 (SR) identified as main sequence stars (asterisks) and 40 by
     radial-velocity selection (filled dots). 
     Those also selected as members in a preliminary analysis are highlighted
     in gray; members find to be doubtful (due to their measured abundances
     or lack of any
     metallicity estimate) are marked with a plus sign.}
   \label{m15_MvG}
\end{figure}

The globular cluster ${\rm M~15}$ is located in the sky at RA=$21^{\rm h}\,
29^{\rm m}\,58.3^{\rm s}$, DEC=$+12^\circ \,10\arcmin \,01\arcsec$
\citep{harris}, and has been extensively studied in the past \citep[e.g.,
][]{sandage, GA}.  SDSS/SEGUE plates $1960$ and $1962$ include observations of
its members. Figure~\ref{m15_sel} shows the distribution of the 526 stars with
available SDSS/SEGUE spectroscopy and $ugriz$ photometry. The central regions
of the clusters are not generally available for spectroscopic observation, due
to fibre placement restrictions in the SDSS spectrographs. This must be borne
in mind when interpreting the results we describe below.

Based on position, we initially select 133 candidate members in the region
$322^\circ. 25 < \alpha < 322^\circ. 75$ and $11^\circ. 90 < \delta <
12^\circ. 40$, as represented by the box shown in Fig.~\ref{m15_sel}.  

The distribution of the 
atmospheric parameters ${\rm [Fe/H]}$ versus $\log~g$ of this sample,
using both the RR and SR models, is shown in Fig.~\ref{m15_MvG}. The stars
clearly fall into two groups, due to false cluster members which we can
plausibly take to be stars projected in front of the cluster from the Galactic
field (generally at higher metallicity), and stars from the globular cluster
itself (lower metallicity).  It is also obvious that, given the apparent
magnitude limits of SDSS/SEGUE, we would not expect to see higher-gravity main
sequence stars that are true cluster members.

To obtain a more clean sample of likely cluster members, we select from the
observed sample using published estimates of radial velocities and
metallicities for the cluster (see Table~\ref{t_GC}).

We first select based on radial velocity; specifically, we retain as
candidates only those stars with $-126~{\rm km~s^{-1}}  < V_R <
-100~{\rm km~s^{-1}}$. This cut preferentially excludes
metal-rich main sequence stars, and results in a remaining sample that
contains 40 candidates with ${\rm [Fe/H]}<
-1.5$ out of a total of 42. 

We define a second sample, now of main sequence stars; namely, the 8 or 7 stars
(for RR/SR respectively) having metal abundance ${\rm [Fe/H]}< -1.5$ and
$\log~g > 3.5$, without any radial velocity selection.
Using the absolute magnitude determination for
late-type dwarfs as a function of SDSS photometry \citep{bilir} 
\begin{equation}
M_g=5.79(g-r)+1.242(r-i)+1.412
\end{equation}
this second sample shows a distance modulus $(m-M)=14.67$, in agreement with
the typical value $(m-M)_{\rm M~15}=14.93$ reported by \citet{sandage}. 

The complete sample of ${\rm M~15}$ cluster members has 46 (RR)/ 45 (SR)
stars. The entire radial velocity selected sample is shown in
Fig.\ref{m15_MvG} with filled circles, while the metal-poor main sequence
stars we suspect are cluster members are shown with asterisks.

${\rm M~15}$ has been previously analysed during the course of the development
of the SSPP. Our
initial sample (of 133 stars) includes 26 of the 35 candidates. Of these, 7
stars which have been rejected on the grounds of their apparently discrepant
estimated abundance, or lack of an estimate at all, are marked with a plus
sign. The 19 stars confirmed as likely members are also identified as part of
our candidate members; we highlight these as light-colour dots in
Fig.~\ref{m15_MvG}.  
Inspection of this figure shows the ${\rm M~15}$ members as a clump of stars,
albeit one which is more clumped in the RR predictions of the 
atmospheric parameters than in the SR predictions of the 
atmospheric parameters.

From the sample of cluster members with consistent metallicities and radial
velocities we obtain a mean metallicity of 
${\rm [Fe/H]}= -2.20\pm 0.11$~dex (RR)/${\rm [Fe/H]}= -2.26\pm 0.26$~dex (SR).
Using just the giants in this sample (i.e., excluding the metal-poor main
sequence stars) we obtain 
${\rm [Fe/H]}= -2.20\pm 0.10$~dex (RR)/${\rm [Fe/H]}= -2.35\pm 0.14$~dex (SR).
These values are in good agreement with previous
determinations in the literature (see Table~\ref{t_GC}).

\begin{figure*}
   \centering
   \resizebox{\hsize}{!}{\includegraphics[angle=-0,width=17cm]
        {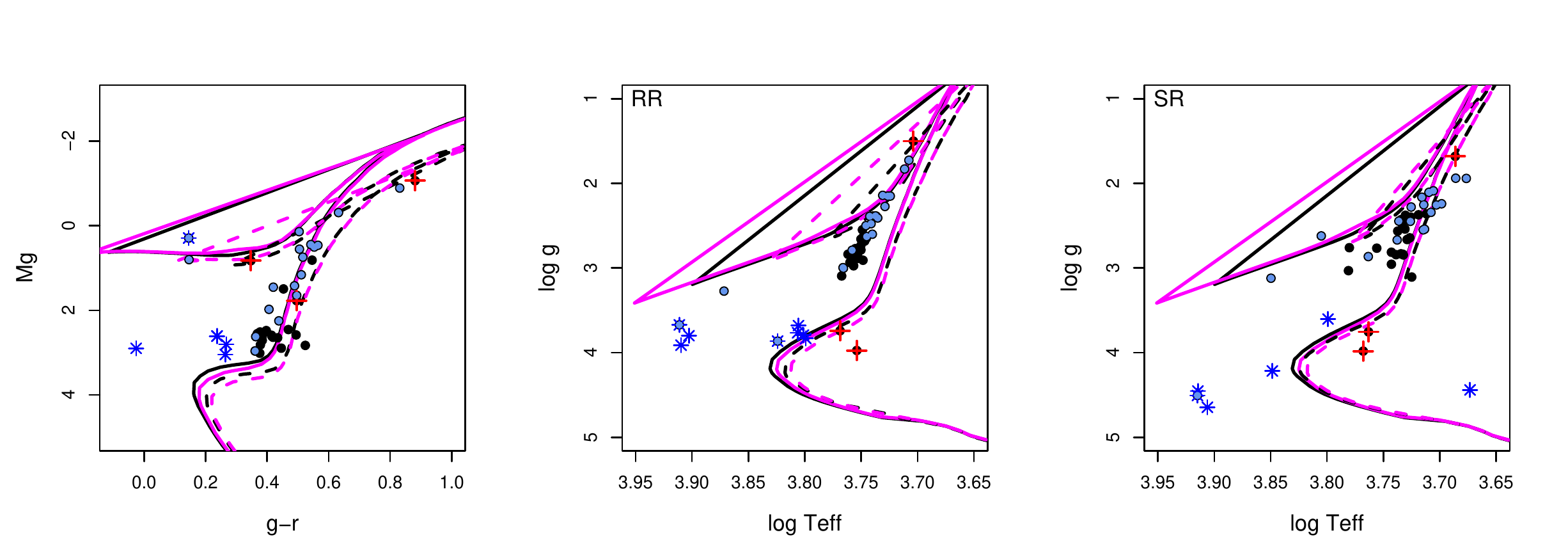}}
   \caption{The left panel shows the colour--magnitude diagram for M~15, and
     the two other panels the 
     distribution of atmospheric parameters $\log~g$ vs. $\log~T_{\rm eff}$
   from the RR (middle) and SR (right) models. 
     Of the stars selected as candidates, the asterisks denote main sequence
     metal-poor stars, the filled dots the members based on radial velocity
     constrain. Confirmed and doubtful members assigned in a preliminary
     analysis are coloured gray and marked as plus sign
     respectively. Overplotted are isochrones from \citet{girardi} 
     with metallicities and ages which bracket the values given in
       Table~\ref{t_GC}, i.e. 
     at $Z=0.0001$ (solid), $Z=0.0004$ (dashed) for ages of $12.59$~Gyr (black)
     and $14.13$~Gyr (gray).} 
   \label{cluster_m15}
\end{figure*}

\subsection{${\rm M~15}$ - Isochrones}\label{sect_GCisochrones}

We now compare our 
atmospheric parameter estimates with theoretical SDSS isochrones from
\citet{girardi}.  We adopt an age of 13.2~Gyr, a metallicity ${\rm [Fe/H]}=
-2.22$~dex, and a distance modulus of 14.93 kpc \citep[e. g., ][]{sandage, GA}.

Figure~\ref{cluster_m15} shows the colour--magnitude and effective
temperature-gravity diagrams for the likely ${\rm M~15}$ members overplotted
with the theoretical isochrones.  These isochrones bracket the candidates
reasonably well in the colour--magnitude diagram, but the distribution
in the atmospheric parameter 
plane shows systematic offsets, in particular for the RR model
estimates. A zero-point offset in either the gravity or temperature
parameterizations (or in the isochrones) would improve the coincidence.  On
the other hand, the RR model clearly yields a tighter distribution in the 
atmospheric parameters.
Thus, if we believe the isochrones, then we can conclude that the RR model
obtains more {\em precise} 
parameter estimates, while the SR model obtains more {\em
  accurate} ones. In fact, if we would attribute the offset due entirely to
gravity, we would have to apply corrections of about 0.60 dex (RR) or 0.25 dex
(SR) to our estimates in order to obtain coincidence with their predicted
location in the 
effective temperature-gravity planes.

\subsection{Other Clusters}

We carried out the same analysis for three additional clusters which have also
been extensively studied in the past, and so have reasonably consistent
determinations of metallicity, age, and distance in the literature (see
Table~\ref{t_GC}). Candidates stars from the globular clusters ${\rm M~13}$
\citep[e.g., ][]{sandage,lupton,sketrone, harris,GA} appear on SEGUE plates
$2174$, $2185$, and $2255$; from the globular cluster ${\rm M~2}$ \citep[e.g.,
][]{harris,lazaro} on SEGUE plate $1961$; and from the open cluster ${\rm
NGC~2420}$ \citep[e.g.; ][]{McClure, smith_v, tianxing} on SEGUE plates $2078$
and $2079$. For each of these, we select likely members following the same
procedures as for the ${\rm M~15}$ analysis (Sect.~\ref{sect_GCselection}) and
compare them with isochrones with parameters based on previous analyses.

\begin{figure*}
   \centering
   \includegraphics[angle=-0,width=17cm]{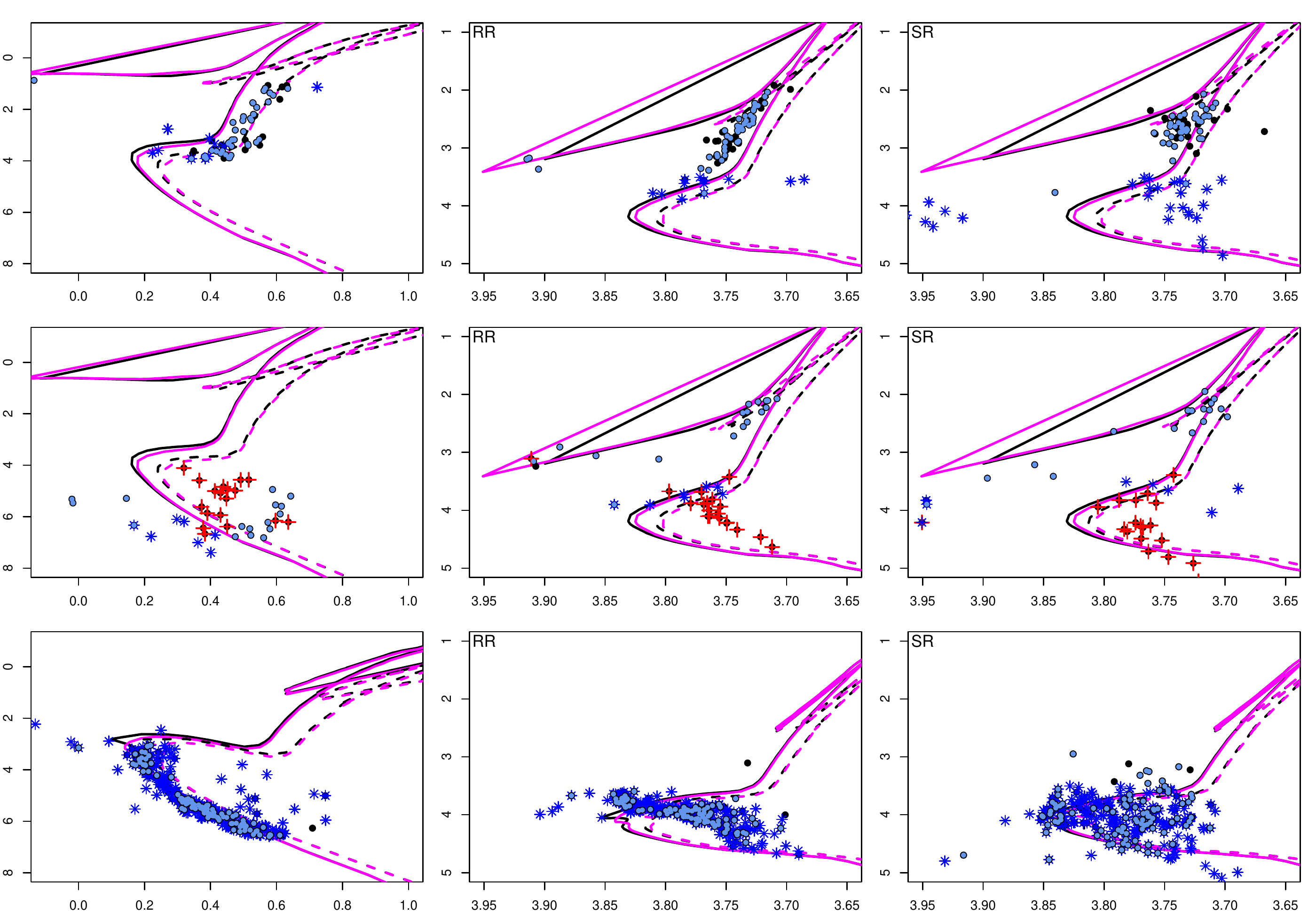}
   \caption{As Fig.~\ref{cluster_m15}. Top: ${\rm M~13}$ (globular cluster)
     candidates. Centre: ${\rm M~2}$ (globular cluster) candidates. Bottom:
     ${\rm NGC~2420}$ (open cluster) candidates. 
     For the globular clusters, isochrones at $Z=0.0001$ (solid),
       $Z=0.0004$ (dashed) for ages of $12.59$~Gyr (black) and $14.13$~Gyr
       (gray); for the open cluster, isochrones at $Z=0.004$ (solid),
       $Z=0.008$ (dashed) for ages of $3.162$~Gyr (black) and $3.548$~Gyr
       (gray).}
   \label{clusters}
\end{figure*}

Figure~\ref{clusters} shows the distribution of the 
atmospheric parameters for expected members of
each cluster in the colour--magnitude and in the ($\log~T_{\rm
  eff}$--$\log~g$) plane, overplotted with the theoretical isochrones selected
to best match each cluster's properties. Inspection of these distributions
confirms our previous conclusions for the case of ${\rm M~15}$ --  (1) there
exists a systematic offset in effective temperature and/or surface gravity
between the estimated 
parameters and those expected from the theoretical isochrones,
and (2) the RR model provides more precise atmospheric parameter
estimates, 
while the SR model provides more accurate ones. 

We are limited by the small number of likely cluster members in some cases,
especially for ${\rm M~2}$, which (so far) appears on only one SEGUE plate. 
However, it seems that this evidence is more clearly visible in the globular
clusters which, as for ${\rm M~15}$, are old and metal poor. 
In the atmospheric parameter plane, the distribution for the open
cluster ${\rm NGC~2420}$ from the SR model looks a bit confusing. It is
plausible that this cluster is too metal-rich 
to obtain good 
atmospheric parameter estimates, as the expected parameters are at the
extreme of the regions covered by the synthetic grid used for training.  Larger
uncertainties are certainly present in this range of metallicity 
(see Tables~\ref{t_Partial1SR}, ~\ref{t_Partial2SR}).   These limitations are
under study at the moment.  

\begin{table*}
  \caption{Globular/Open Clusters, literature values. 
    The selection constraints applied for identification of likely members are labeled with $*$.
}
  \hspace{21 cm}
  \label{t_GC}
  \begin{center}
    \leavevmode
        \begin{tabular}[h]{lccccccrr}
        \hline\hline \\[-5pt]
    Cluster & (RA, DEC) & RA$^*$ & DEC$^*$ & ${\rm [Fe/H]}$& age & m-M & RV & RV$^*$\\
            & &(degree) & (degree) & (dex) & (Gyr) & (kpc) & ${\rm
              (km~s^{-1})}$ & ${\rm
              (km~s^{-1})}$ \\[+5pt]
        \hline \\[-5pt]
    ${\rm M~15}$ &$21^{\rm h}\,29^{\rm m}\,58.3^{\rm s}$, $+12^\circ
    \,10\arcmin\,01\arcsec$ & $322.25,322.75$ & $11.90,12.40$ & $-2.22$ &
    $13.2$ & $14.93$ & $-110$ & $-126,-100$ \\
    ${\rm M~13}$ &$16^{\rm h}\,41^{\rm m}\, 41.5^{\rm s}$, $+36^\circ
    \,27\arcmin\,37\arcsec$ & $250.00,250.90$ & $36.10,36.90$ & $-1.70$ &
    $12.7$ & $14.07$ & $-250$ & $-262,-243$\\
    ${\rm M~2}$  &$21^{\rm h}\,33^{\rm m}\, 29.3^{\rm s}$, $-00^\circ \,
    49\arcmin\,23\arcsec$ & $323.10,323.60$ & $-1.05,-0.60$ & $-1.53$ & $13.0$
    & $10.49$ & $0$ & $-20,20$\\
    ${\rm NGC~2420}$ &$07^{\rm h}\,38^{\rm m}\, 24.0^{\rm s}$, $+21^\circ \,
    34\arcmin\,27\arcsec$ & $114.40,115.10$ & $21.20,22.10$ & $-0.50$ & $3/4$
    & $11.40$ & $73$ & $50,86$\\[+5pt]
\hline \\[-5pt]
       \end{tabular}
  \end{center}
\end{table*}
\section{Summary and conclusions}\label{concluding}

We have developed models to estimate the three primary stellar atmospheric 
parameters (
$T_{\rm eff}$, $\log ~g$, and ${\rm [Fe/H]}$) from 
SDSS/SEGUE spectra. These models produce self-consistent parameter 
estimates and can be implemented into an automated data processing 
pipeline.  Our models rely on an initial configuration (or ``training'') 
phase, which for one of the models (RR) uses pre-classified observed data, 
for the other (SR) synthetic spectra selected by the user.  Both are 
flexible, in that new models can easily be introduced by changing the set 
of training templates.

Both models are nonlinear, regularized regression models. The RR model 
uses an initial PCA compression of the data to reduce the dimensionality 
(from 3818 to 50), thus producing a more robust (and precise) parametrizer 
(which reduces the dimensionality further to 3, i.e., the three 
atmospheric parameters).   
They are also rapid, requiring of the order of one millisecond per star on a 
single, modest CPU.

The RR model has the advantage that exactly the same type of data are used 
in the training and application phases, thus eliminating the issue of 
discrepancies in the flux calibration or cosmic variance of the two 
samples. Of course, this requires an independent 
estimation method (``basis parameterizer'') to parametrize the training
templates (which itself must use synthetic models at some level).  Our
regression model then automates and -- more importantly -- generalizes this
basis parameterizer. Indeed, the basis parameterizer may even comprise
multiple algorithms, perhaps operating over different 
parameters ranges or used in a voting 
system to estimate atmospheric parameters. 
This is true in the present case, where the basis parameterizer comes from a
preliminary version of the SDSS/SEGUE Spectroscopic Parameter Pipeline
\citep[SSPP; ][]{beers06, lee07}. 

In contrast, our SR model is trained directly on synthetic spectra, 
dispensing with the need for a basis parameterizer. For best results these 
training data should have noise properties similar to the observed data 
(which improves the regularization). We therefore implemented different models 
for different SNR ranges.  PCA is again used for data compression, except 
for the surface gravity parameter $\log ~g$, where better results were obtained
using a subset of spectral features known to be most sensitive to this
parameter.

For each 
atmospheric parameter, the accuracy of our predictions with respect to
previous estimates (SSPP) are $T_{\rm eff}$ to $170/170$~K, $\log~g$ to
$0.36/0.45$~dex and ${\rm [Fe/H]}$ to $0.19/0.26$~dex for methods RR and SR
respectively.  Consistency between the two approaches is on order of $150$~K
in $T_{\rm eff}$, $0.35$~dex in $\log~g$, and $0.22$~dex in ${\rm
  [Fe/H]}$. Some discrepancies are probably due to the different Kurucz models
adopted in our SR model and in some of the methods employed in the SSPP. 

As a test of our model predictions, we estimated 
atmospheric parameters for globular/open cluster 
members and compared these to theoretical isochrones. We found that RR 
gives more {\em precise} 
parameter estimates (stars show smaller scatter) whereas 
SR gives more {\em accurate} ones (stars show smaller offset, or bias). We 
can use this information to improve the 
parameter calibration of the basis 
parametrizers or the pre-processing of the synthetic spectra. We have also 
used our models to estimate 
atmospheric parameters for 89\,600 SEGUE and 194\,172 SDSS (DR-5) 
stellar spectra, which are being used for further scientific investigations.

We found that the inclusion of the four SDSS photometric colours improves 
the precision of parameter estimation significantly, but
this will only work for zero (or very low) extinction regions.  In principle,
our models can be extended to predict extinction (by inclusion of its variance
in the training set), allowing us to then use both photometry and spectroscopy
to predict 
atmospheric parameters along significantly reddened lines of sight. 

Our RR model has already been successfully integrated into the SSPP.  The 
SR will undergo further refinement with improved synthetic spectra.  In 
particular, models with more molecules included in the linelists
will improve the representation of cool stars.  An extension to hotter stars
will make the model more widely applicable (at present such stars can be
filtered out via the PCA reconsuction error).  Looking further ahead, the SR
approach will form the basis for 
 atmospheric parameter estimation from the very low resolution
spectrophotometry (R$\simeq$12--40) to be obtained with Gaia (albeit using a
more sophisticated and knowledge-based approach to regression, which also
includes the accurate parallaxes and high-precision photometry from Gaia).  
Our pattern recognition approach is probably indispensable in such an
application, because the low resolution and spectral purity of the
spectrophotometry prevent the definition of traditional indices.

\clearpage
\begin{table*}
  \caption{RR: Partial results. We list the mean $\mu$ and the
    corresponding standard deviation $\sigma$ of the difference
  Committee-SSPP for each of the different stellar types and parameter
  ranges.}
  \hspace{21 cm}
  \label{t_Partial1RR}
  \begin{center}
    \leavevmode
        \begin{tabular}[h]{ccrrrrrrrrr}
        \hline\hline \\[-5pt]
        ${\rm [Fe/H]}$ & $\log~T_{\rm eff}$ & 
        $\mu_{\log~T_{\rm eff}}$&$\sigma_{\log~T_{\rm eff}}$&$E_{\log~T_{\rm eff}}$ &
        $\mu_{\log~g}$&$\sigma_{\log~g}$&$E_{\log~g}$&
        $\mu_{\rm [Fe/H]}$&$\sigma_{\rm [Fe/H]}$& $E_{\rm [Fe/H]}$\\[+5pt]
        \hline \\[-5pt]
$<-1.5$ & $<3.70$ & $0.0155$ & $0.0166$ & $0.0172$ & $0.1879$ & $0.4448$ & $0.3724$ & $0.2673$ & $0.2733$ & $0.3056$\\
$<-1.5$ & $3.70,3.75$ & $0.0061$ & $0.0134$ & $0.0105$ & $0.1214$ & $0.3939$ & $0.3075$ & $0.0817$ & $0.2072$ & $0.1464$\\
$<-1.5$ & $3.75,3.80$ & $0.0011$ & $0.0097$ & $0.0073$ & $0.1284$ & $0.3463$ & $0.2823$ & $0.0619$ & $0.1531$ & $0.1266$\\
$<-1.5$ & $3.80,3.85$ & $-0.0032$ & $0.0107$ & $0.0082$ & $-0.1141$ & $0.4439$ & $0.3606$ & $0.0305$ & $0.2048$ & $0.1484$\\
$<-1.5$ & $3.85,3.875$ & $0.0146$ & $0.0128$ & $0.0167$ & $0.1617$ & $0.5463$ & $0.4117$ & $0.1070$ & $0.3837$ & $0.2882$\\
$<-1.5$ & $>3.875$ & $-0.0044$ & $0.0230$ & $0.0173$ & $0.0007$ & $0.3487$ & $0.2412$ & $0.2860$ &$0.4164$ & $0.3875$\\[+5pt]
\hline \\[-5pt]
$>-1.5$ & $<3.70$ & $0.0087$ & $0.0129$ & $0.0118$ & $0.0941$ & $0.3696$ & $0.2882$ & $0.0295$ & $0.2150$ & $0.1658$\\
$>-1.5$ & $3.70,3.75$ & $0.0010$ & $0.0112$ & $0.0083$ & $0.0206$ & $0.3187$ & $0.2466$ & $0.0078$ & $0.1403$ & $0.1044$\\
$>-1.5$ & $3.75,3.80$ & $-0.0034$ & $0.0107$ & $0.0084$ & $-0.0765$ & $0.3051$ & $0.2458$ & $-0.0359$ & $0.1469$ & $0.1139$\\
$>-1.5$ & $3.80,3.85$ & $-0.0052$ & $0.0127$ & $0.0097$ & $-0.0698$ & $0.4512$ & $0.3462$ & $-0.0923$ & $0.2307$ & $0.1904$\\
$>-1.5$ & $3.85,3.875$ & $0.0022$ & $0.0105$ & $0.0081$ & $-0.0786$ & $0.5434$ & $0.4127$ & $-0.0484$ & $0.2511$ & $0.1898$\\
$>-1.5$ & $>3.875$ & $-0.0119$ & $0.0189$ & $0.0151$ & $-0.0361$ & $0.4226$ & $0.3085$ & $-0.2144$ & $0.3799$ & $0.3226$\\[+5pt]
\hline\hline \\[-5pt]
        $\log~g$ & $\log~T_{\rm eff}$ & 
        $\mu_{\log~T_{\rm eff}}$&$\sigma_{\log~T_{\rm eff}}$&$E_{\log~T_{\rm eff}}$ &
        $\mu_{\log~g}$&$\sigma_{\log~g}$&$E_{\log~g}$&
        $\mu_{\rm [Fe/H]}$&$\sigma_{\rm [Fe/H]}$& $E_{\rm [Fe/H]}$\\[+5pt]
        \hline \\[-5pt]
$<3.5$ & $<3.70$ & $0.0193$ & $0.0164$ & $0.4302$ & $0.3427$ & $0.4407$ & $0.4302$ & $0.1685$ & $0.2540$ & $0.2347$\\
$<3.5$ & $3.70,3.75$ & $0.0060$ & $0.0150$ & $0.3577$ & $0.2045$ & $0.4069$ & $0.3577$ & $0.0518$ & $0.1950$ & $0.1454$\\
$<3.5$ & $3.75,3.80$ & $0.0034$ & $0.0112$ & $0.4115$ & $0.3747$ & $0.3202$ & $0.4115$ & $0.0687$ & $0.1719$ & $0.1401$\\
$<3.5$ & $3.80,3.85$ & $0.0018$ & $0.0128$ & $0.6220$ & $0.6182$ & $0.3584$ & $0.6220$ & $0.0516$ & $0.2630$ & $0.1946$\\
$<3.5$ & $3.85,3.875$ & $0.0122$ & $0.0139$ & $0.5691$ & $0.5357$ & $0.5020$ & $0.5691$ & $0.1258$ & $0.2992$ & $0.2347$\\
$<3.5$ & $>3.875$ & $-0.0078$ & $0.0240$ & $0.2639$ & $0.1860$ & $0.3137$ & $0.2639$ & $0.0662$ & $0.4635$ & $0.3435$\\[+5pt]
\hline \\[-5pt]
$>3.5$ & $<3.70$ & $0.0069$ & $0.0114$ & $0.0105$ & $0.0438$ & $0.3373$ & $0.2644$ & $0.0294$ & $0.2226$ & $0.1692$\\
$>3.5$ & $3.70,3.75$ & $0.0001$ & $0.0099$ & $0.0075$ & $-0.0207$ & $0.2805$ & $0.2216$ & $0.0024$ & $0.1287$ & $0.0968$\\
$>3.5$ & $3.75,3.80$ & $-0.0035$ & $0.0100$ & $0.0080$ & $-0.1172$ & $0.2509$ & $0.2188$ & $-0.0300$ & $0.1434$ & $0.1116$\\
$>3.5$ & $3.80,3.85$ & $-0.0054$ & $0.0116$ & $0.0091$ & $-0.2033$ & $0.3408$ & $0.3051$ & $-0.0641$ & $0.2185$ & $0.1717$\\
$>3.5$ & $3.85,3.875$ & $0.0019$ & $0.0099$ & $0.0076$ & $-0.2488$ & $0.3877$ & $0.3497$ & $-0.0700$ & $0.2675$ & $0.2005$\\
$>3.5$ & $>3.875$ & $-0.0092$ & $0.0190$ & $0.0146$ & $-0.1452$ & $0.3797$ & $0.2874$ & $-0.0367$ & $0.4618$ & $0.3521$\\[+5pt]
\hline \\[-5pt]
       \end{tabular}
  \end{center}
\end{table*}

\begin{table*}
  \caption{RR: Partial results. We list the mean $\mu$ and the
    corresponding standard deviation $\sigma$ of the difference
  Committee-SSPP for each of the different stellar temperatures and metallicity
  ranges.}
  \hspace{21 cm}
  \label{t_Partial2RR}
  \begin{center}
    \leavevmode
        \begin{tabular}[h]{ccrrr}
        \hline\hline \\[-5pt]
        $T_{\rm eff}$ & ${\rm [Fe/H]}$ & $\mu_{\rm [Fe/H]}$& $\sigma_{\rm [Fe/H]}$& $E_{\rm [Fe/H]}$\\[+5pt]
        \hline \\[-5pt]
$< 4500$    &    $< -2.5$  & $0.9062$ & $0.8550$ & $0.9062$\\
$< 4500$    & $-2.5,-2.0$  & $0.2322$ & $0.1414$ & $0.2322$\\
$< 4500$    & $-2.0,-1.5$  & $0.3464$ & $0.3664$ & $0.4503$\\
$< 4500$    & $-1.5,-1.0$  & $0.1196$ & $0.1779$ & $0.1667$\\
$< 4500$    & $-1.0,-0.5$  &$-0.1419$ & $0.1634$ & $0.1732$\\
$< 4500$    &    $ > -0.5$  &$-0.4616$ & $0.1322$ & $0.4616$\\[+5pt]
\hline \\[-5pt]
$4500,6500$ &    $< -2.5$  & $0.1183$ & $0.1770$ & $0.1574$\\
$4500,6500$ & $-2.5,-2.0$  &$-0.0043$ & $0.1608$ & $0.1144$\\
$4500,6500$ & $-2.0,-1.5$  &$-0.0133$ & $0.1636$ & $0.1219$\\
$4500,6500$ & $-1.5,-1.0$  &$-0.0435$ & $0.1683$ & $0.1303$\\
$4500,6500$ & $-1.0,-0.5$  &$-0.0294$ & $0.1291$ & $0.0999$\\
$4500,6500$ &     $> -0.5$  &$-0.1416$ & $0.1207$ & $0.1456$\\[+5pt]
\hline \\[-5pt]

       \end{tabular}
  \end{center}
\end{table*}

\clearpage
\begin{table*}
  \caption{SR: Partial results. We list the mean $\mu$ and the
    corresponding standard deviation $\sigma$ of the difference
  Committee-SSPP for each of the different stellar types and parameter
  ranges.}
  \hspace{21 cm}
  \label{t_Partial1SR}
  \begin{center}
    \leavevmode
        \begin{tabular}[h]{ccrrrrrrrrr}
        \hline\hline \\[-5pt]
        ${\rm [Fe/H]}$ & $\log~T_{\rm eff}$ & 
        $\mu_{\log~T_{\rm eff}}$&$\sigma_{\log~T_{\rm eff}}$&$E_{\log~T_{\rm eff}}$ &
        $\mu_{\log~g}$&$\sigma_{\log~g}$&$E_{\log~g}$&
        $\mu_{\rm [Fe/H]}$&$\sigma_{\rm [Fe/H]}$& $E_{\rm [Fe/H]}$\\[+5pt]
        \hline \\[-5pt]
$<-1.5$ & $<3.70$ & $-0.0076$ & $0.0272$ & $0.0212$ & $0.4070$ & $0.6537$ & $0.6097$ & $-0.4354$ & $0.4470$ & $0.4902$\\
$<-1.5$ & $3.70,3.75$ & $-0.0105$ & $0.0143$ & $0.0143$ & $-0.0665$ & $0.4660$ & $0.3627$ & $-0.3936$ & $0.3146$ & $0.4099$\\
$<-1.5$ & $3.75,3.80$ & $-0.0107$ & $0.0129$ & $0.0133$ & $-0.4136$ & $0.4600$ & $0.5013$ & $-0.1549$ & $0.3294$ & $0.2454$\\
$<-1.5$ & $3.80,3.85$ & $-0.0047$ & $0.0117$ & $0.0098$ & $-0.5609$ & $0.5208$ & $0.6488$ & $-0.1458$ & $0.3791$ & $0.2873$\\
$<-1.5$ & $3.85,3.875$ & $0.0051$ & $0.0147$ & $0.0099$ & $-0.0387$ & $0.6575$ & $0.4753$ & $-0.0771$ & $0.4216$ & $0.3410$\\
$<-1.5$ & $>3.875$ & $0.0171$ & $0.0230$ & $0.0228$ & $0.6129$ & $0.5321$ & $0.6973$ & $0.1750$ & $0.5635$ & $0.4622$\\[+5pt]
\hline \\[-5pt]
$>-1.5$ & $<3.70$ & $0.0288$ & $0.0254$ & $0.0338$ & $0.1509$ & $0.5730$ & $0.4750$ & $0.1275$ & $0.3951$ & $0.3267$\\
$>-1.5$ & $3.70,3.75$ & $0.0030$ & $0.0164$ & $0.0112$ & $-0.0213$ & $0.5409$ & $0.4298$ & $0.0376$ & $0.3322$ & $0.2593$\\
$>-1.5$ & $3.75,3.80$ & $0.0054$ & $0.0160$ & $0.0131$ & $-0.1007$ & $0.4923$ & $0.3937$ & $0.0176$ & $0.2545$ & $0.1955$\\
$>-1.5$ & $3.80,3.85$ & $0.0045$ & $0.0140$ & $0.0099$ & $-0.1443$ & $0.5179$ & $0.4154$ & $0.0415$ & $0.2876$ & $0.2261$\\
$>-1.5$ & $3.85,3.875$ & $0.0136$ & $0.0119$ & $0.0142$ & $0.0482$ & $0.5324$ & $0.4087$ & $0.0744$ & $0.3325$ & $0.2408$\\
$>-1.5$ & $>3.875$ & $0.0310$ & $0.0200$ & $0.0320$ & $0.6196$ & $0.6062$ & $0.7288$ & $0.2546$ & $0.4465$ & $0.3759$\\[+5pt]
\hline\hline \\[-5pt]
        $\log~g$ & $\log~T_{\rm eff}$ & 
        $\mu_{\log~T_{\rm eff}}$&$\sigma_{\log~T_{\rm eff}}$&$E_{\log~T_{\rm eff}}$ &
        $\mu_{\log~g}$&$\sigma_{\log~g}$&$E_{\log~g}$&
        $\mu_{\rm [Fe/H]}$&$\sigma_{\rm [Fe/H]}$& $E_{\rm [Fe/H]}$\\[+5pt]
        \hline \\[-5pt]
$3.5$ & $<3.70$ & $-0.0099$ & $0.0278$ & $0.4282$ & $0.0085$ & $0.5627$ & $0.4282$ & $-0.3130$ & $0.4795$ & $0.3967$\\
$3.5$ & $3.70,3.75$ & $-0.0065$ & $0.0150$ & $0.4088$ & $-0.2002$ & $0.4827$ & $0.4088$ & $-0.2663$ & $0.2919$ & $0.3121$\\
$3.5$ & $3.75,3.80$ & $-0.0068$ & $0.0160$ & $0.5700$ & $-0.5161$ & $0.4468$ & $0.5700$ & $-0.1164$ & $0.2867$ & $0.2257$\\
$3.5$ & $3.80,3.85$ & $-0.0015$ & $0.0127$ & $0.6803$ & $-0.6085$ & $0.5020$ & $0.6804$ & $-0.1003$ & $0.3585$ & $0.2766$\\
$3.5$ & $3.85,3.875$ & $0.0154$ & $0.0172$ & $0.5740$ & $-0.2528$ & $0.6760$ & $0.5740$ & $0.0592$ & $0.4546$ & $0.3449$\\
$3.5$ & $>3.875$ & $0.0359$ & $0.0290$ & $0.5496$ & $0.1515$ & $0.6719$ & $0.5496$ & $0.3883$ & $0.4792$ & $0.4839$\\[+5pt]
\hline \\[-5pt]
$3.5$ & $<3.70$ & $0.0246$ & $0.0275$ & $0.0322$ & $0.3073$ & $0.6112$ & $0.5486$ & $0.0149$ & $0.4690$ & $0.3772$\\
$3.5$ & $3.70,3.75$ & $0.0033$ & $0.0167$ & $0.0120$ & $0.0297$ & $0.5323$ & $0.4223$ & $0.0471$ & $0.3540$ & $0.2741$\\
$3.5$ & $3.75,3.80$ & $0.0057$ & $0.0156$ & $0.0130$ & $0.0039$ & $0.4320$ & $0.3393$ & $0.0229$ & $0.2728$ & $0.1987$\\
$3.5$ & $3.80,3.85$ & $0.0034$ & $0.0144$ & $0.0101$ & $-0.0841$ & $0.4873$ & $0.3789$ & $0.0301$ & $0.3038$ & $0.2275$\\
$3.5$ & $3.85,3.875$ & $0.0119$ & $0.0117$ & $0.0130$ & $0.0866$ & $0.5113$ & $0.3907$ & $0.0520$ & $0.3288$ & $0.2391$\\
$3.5$ & $>3.875$ & $0.0272$ & $0.0207$ & $0.0297$ & $0.6556$ & $0.5666$ & $0.7354$ & $0.2241$ & $0.4747$ & $0.3885$\\[+5pt]
\hline \\[-5pt]
       \end{tabular}
  \end{center}
\end{table*}

\begin{table*}
  \caption{SR: Partial results. We list the mean $\mu$ and the
    corresponding standard deviation $\sigma$ of the difference
  Committee-SSPP for each of the different stellar temperatures and metallicity
  ranges.}
  \hspace{21 cm}
  \label{t_Partial2SR}
  \begin{center}
    \leavevmode
        \begin{tabular}[h]{ccrrr}
        \hline\hline \\[-5pt]
        $T_{\rm eff}$ & ${\rm [Fe/H]}$ & $\mu_{\rm [Fe/H]}$& $\sigma_{\rm [Fe/H]}$& $E_{\rm [Fe/H]}$\\[+5pt]
        \hline \\[-5pt]
$< 4500$    &    $< -2.5$  & $-0.9158$ &$0.6332$& $0.9158$\\
$< 4500$    & $-2.5,-2.0$  & $-0.2961$ &$0.4686$& $0.4342$\\
$< 4500$    & $-2.0,-1.5$  & $ 0.2013$ &$0.6026$& $0.4545$\\
$< 4500$    & $-1.5,-1.0$  & $ 0.2022$ &$0.4808$& $0.3597$\\
$< 4500$    & $-1.0,-0.5$  & $-0.4289$ &-& $0.4289$\\
$< 4500$    &    $ > -0.5$  &        -&-&      -\\[+5pt]
\hline \\[-5pt]
$4500,6500$ &    $< -2.5$  & $-0.6448$ & $0.5575$& $0.6678$\\
$4500,6500$ & $-2.5,-2.0$  & $-0.2419$ & $0.3023$& $0.2879$\\
$4500,6500$ & $-2.0,-1.5$  & $-0.1947$ & $0.3054$& $0.2832$\\
$4500,6500$ & $-1.5,-1.0$  & $-0.1968$ & $0.2580$& $0.2673$\\
$4500,6500$ & $-1.0,-0.5$  & $ 0.0007$ & $0.2260$& $0.1700$\\
$4500,6500$ &     $> -0.5$  & $ 0.3192$ & $0.2566$& $0.3300$\\[+5pt]
\hline \\[-5pt]

       \end{tabular}
  \end{center}
\end{table*}

\begin{acknowledgements}
This work was partly funded by a DFG Emmy-Noether Nachwuchsgruppe grant
to C.A.L. Bailer-Jones.  Y.S. Lee, T.C. Beers, and T. Sivarani acknowledge
partial support for this work from  grant AST 04-06784, as well as 
from grant PHY 02-16783: Physics Frontiers Center/Joint Institute for Nuclear
Astrophysics (JINA), both awarded by the U.S. National Science Foundation.

We wish to thank the referee, Norbert Christlieb, for a careful reading
  of this manuscript and for his useful remarks.

Funding for the SDSS and SDSS-II has been provided by the Alfred P. Sloan
Foundation, the Participating Institutions, the National Science Foundation,
the U.S. Department of Energy, the National Aeronautics and Space
Administration, the Japanese Monbukagakusho, the Max Planck Society, and the
Higher Education Funding Council for England. The SDSS Web Site is
http://www.sdss.org/.

    The SDSS is managed by the Astrophysical Research Consortium for the
    Participating Institutions. The Participating Institutions are the
    American Museum of Natural History, Astrophysical Institute Potsdam,
    University of Basel, University of Cambridge, Case Western Reserve
    University, University of Chicago, Drexel University, Fermilab, the
    Institute for Advanced Study, the Japan Participation Group, Johns Hopkins
    University, the Joint Institute for Nuclear Astrophysics, the Kavli
    Institute for Particle Astrophysics and Cosmology, the Korean Scientist
    Group, the Chinese Academy of Sciences (LAMOST), Los Alamos National
    Laboratory, the Max-Planck-Institute for Astronomy (MPIA), the
    Max-Planck-Institute for Astrophysics (MPA), New Mexico State University,
    Ohio State University, University of Pittsburgh, University of Portsmouth,
    Princeton University, the United States Naval Observatory, and the
    University of Washington.

\end{acknowledgements}


\end{document}